\documentclass[preprint,12pt]{elsarticle}

\newcommand {\QED}{\hfill$\hspace*{\fill}\rule{1ex}{1ex}$\ \par\vskip .1in}

	\newcommand{\CD}{{\cal D}}
	\newcommand{\DO}{{\cal U}}
	\newcommand{\CU}{{\cal U}}
	\newcommand{\CM}{{\cal M}}

	\newcommand{\CR}{{\cal R}}
	
	\newcommand{\CN}{{\cal N}}

	\newcommand{\CV}{{\cal V}}

	\usepackage[linesnumbered, noend, ruled, vlined]{algorithm2e}
\usepackage{amssymb}
\usepackage{amsmath}
\usepackage{amsthm}

\newtheorem{example}{Example}
\newtheorem{definition}{Definition}%
\newtheorem{lemma}{Lemma}
\newtheorem{proposition}{Proposition}

\newtheorem{theorem}{Theorem}
\newdefinition{rmk}{Remark}
\newproof{pf}{Proof}
\newproof{pot}{Proof of Theorem \ref{thm2}}

\journal{Information Systems}

\begin{document}

\begin{frontmatter}

\title{Evaluating Continuous Basic Graph Patterns over Dynamic Link Data Graphs}



\author{Manolis Gergatsoulis} 
\author{Matthew Damigos} 

\affiliation{organization={Laboratory on Digital Libraries and Electronic Publishing,\\ Department of Archives, Library Science and Museology, \\ Ionian University},
            addressline={Ioannou Theotoki 72},
            city={Corfu},
            postcode={49100},
           country={Greece}}

\begin{abstract}
In this paper, we investigate the problem of evaluating Basic Graph Patterns (BGP, for short, a subclass of SPARQL queries) over dynamic Linked Data graphs; i.e., Linked Data graphs that are continuously updated. We consider a setting where the updates are continuously received through a stream of messages and support both insertions and deletions of triples (updates are straightforwardly handled as a combination of deletions and insertions). In this context, we propose a set of in-memory algorithms minimizing the cached data to efficiently and continuously answer BGP queries.  The queries are typically submitted into a system and continuously result in the delta answers while the update messages are processed.
To efficiently and continuously evaluate the submitted query over the streaming data, as well as to minimize the amount of cached data, we propose an approach where the submitted query is decomposed into simpler subqueries and the query evaluation is achieved by combining the intermediate answers of the subqueries. Using this approach, the proposed algorithms compute the delta answers of a BGP query in polynomial time and space. Note that for certain subclasses of BGP queries, we show that the evaluation can be achieved in constant or linear time and space.
Consolidating all the historical delta answers, the algorithms ensure that the answer to each query is constructed at any given time.
\end{abstract}

\begin{keyword}
 Dynamic Linked Data,  BGP Queries,  Question Answering over Linked Data
\end{keyword}

\end{frontmatter}


\section{Introduction}
\label{sec:intro}
	
The dynamic graphs describe graphs that are continuously modified over time, usually, through a stream of edge updates (insertions and deletions). Representative examples include social graphs \cite{balduini2015citysensing}, traffic and transportation networks \cite{lecue2012capturing}, \cite{tallevi2013real}, financial transaction networks \cite{schneider2012microsecond}, and sensor networks \cite{sheth2008semantic}. When the graph data is rapidly updated, conventional graph management systems are not able to handle high velocity data in a reasonable time. In such cases, real-time query answering becomes a major challenge.
	
RDF data model and Linked Data paradigm are widely used to structure and publish  semantically-enhanced data. The last decades, both private and public organizations have been more and more following this approach to disseminate their data. Due to the emergence and spread of IoT, such an approach attracted further attention from both industry and research communities. To query such type of data, SPARQL is a standard query language that is widely used.
	
Querying streaming Linked Data has been extensively investigated in the literature, where the majority of the related work \cite{dell2017stream,margara2014streaming} focus on investigating frameworks for efficiently querying streaming data; mainly focusing on defining certain operators for querying and reasoning data in sliding windows (i.e., predefined portion of the streaming data). In this work, we mainly focus on efficiently handling the continuously updated Linked Data graph (i.e., Dynamic Link Data Graphs). In particular, we consider a setting where the updates are continuously received through a stream of messages and support both insertions and deletions of triples (updates are straightforwardly handled as a combination of deletions and insertions). In this context, we propose a set of in-memory algorithms minimizing the cached data for efficiently and continuously answering Basic Graph Patterns (BGP, for short, a subclass of SPARQL queries). The queries are typically submitted into a system and continuously result the delta answers while the update messages are processed. The evaluation approach is based on applying an effective decomposition of the given queries in order to both improve the performance of query evaluation and minimize the cached data.
To optimize the query evaluation process, we also investigate subclasses of BGP queries and show that for certain subclasses, the overall processing time can be significantly improved.
Consolidating all the historical delta answers, the algorithms ensure that the answer of each query is constructed at any given time.
	
The paper is structured as follows. Section~\ref{sec:related} presents the related work.
The main concepts and definitions used throughout this work are formally presented in Section~\ref{sec:prelim}.
Section~\ref{sec:special} focuses on defining the main subclasses of BGP queries used throughout the paper. The formal definition of the problem, along with the relevant setting, are discussed in Section~\ref{sec:continuous}. Section~\ref{sec:answer-queries} includes the main contributions of this work; i.e., the query answering algorithms applied over the dynamic Linked Data graphs.
In Section~\ref{subsec:generic}, we present a generic approach for continuously evaluating BGP queries, while individual subsections are dedicated to presenting optimized algorithms for selected subclasses of BGP queries (i.e., ground queries in Section~\ref{subsec:ground}, simple var-centric star queries in Section~\ref{subsec:var-centric},  loosely-connected queries in Section~\ref{subsec:loosely-connected}, and var-connected queries in Section~\ref{sec:eval-connected-var-queries}).


\section{Related work}
\label{sec:related}

The problem of querying and detecting graph patterns over streaming RDF and Linked Data has been extensively investigated in the literature \cite{bonte2023streaming, dell2017stream,margara2014streaming}. In this context, there have been proposed and analyzed multiple settings, such as Data Stream Management Systems (DSMS - e.g., \cite{BarbieriBCVG10,BarbieriBCG10,le2011native,jesper2011high,bolles2008streaming}), and Complex Event Processing Systems (CEP - e.g., \cite{GroppeGKL07,anicic2011ep,roffia2018dynamic,PhamAM19,gillani2016continuous}).

In DSMSs, the streaming data is mainly represented by relational messages, and the queries are translated into plans of relational operators. Representative DSMS systems are the following: C-SPARQL \cite{BarbieriBCVG10,BarbieriBCG10}, CQELS \cite{le2011native}, C-SPARQL on S4 \cite{jesper2011high} and Streaming-SPARQL \cite{bolles2008streaming}. Here, it's worth noting the Strider~\cite{ren2017strider}, which proposes a distributed RDF Stream Processing Engine that is built on top of Apache Kafka and Spark.
The majority of the DSMS approaches use sliding windows to limit the data and focus on proposing a framework for querying the streaming data included in the window. In addition, they mainly use relational-like operators, which are evaluated over the data included in the window. Note that this type of systems mainly focuses on querying the data into windows and does not focus on setting up temporal operators.

In a similar context, RSP4J~\cite{tommasini2021rsp4j}, a configurable RSPQL (RDF Stream Processing Query Language) API and engine, accesses streamed RDF data via one or more time-based windows. At each evaluation time, only the triples currently within the window(s) are considered from the incoming stream(s). These windowed RDF triples are then combined with any static background data (if present) to form the time-varying dataset on which the SPARQL-like query is evaluated.

On the other hand, CEP systems follow a different approach~\cite{roffia2018dynamic, anicic2011ep, PhamAM19}. These approaches handle the streaming data as an infinite sequence of events that are used to compute composite events. EP-SPARQL \cite{anicic2011ep} is a representative system of this approach, which defines a language for event processing and reasoning. It also supports windowing operators as well as temporal operators. C-ASP \cite{PhamAM19} is a hybrid approach combining the windowing mechanism and relational operators from DSMS systems and the rule-based and event-based approach of CEP systems. In~\cite{roffia2018dynamic}, the authors extend this approach to a decentralized setting and propose a Web-based software architecture, named SEPA (SPARQL Event Processing Architecture), which enables the development of distributed Web of Data applications. Although update operations are considered, the authors do not investigate in detail how the SPARQL query evaluation is performed.  Table~\ref{table:comparison-table} summarizes the main differences between DSMS and CEP systems, as well as the setting investigated in this paper.

\begin{table}[htbp]\footnotesize
\centering
\begin{tabular}{|p{0.15\linewidth}|p{0.24\linewidth}|p{0.24\linewidth}|p{0.24\linewidth}|}
\hline
\multicolumn{ 1}{|c|}{\textbf{Aspect}} & \multicolumn{ 1}{|c|}{\textbf{DSMS}} & \multicolumn{ 1}{|c|}{\textbf{CEP}} & \multicolumn{ 1}{|c|}{\textbf{Dynamic LDG}} \\ \hline

Handling of Streaming Data & Representing streaming data primarily as relational tuples, processing queries using relational operators over sliding windows. & Streaming data as sequences of events, focusing on event pattern detection and aggregation to infer high-level composite events. & Dynamic graphs where updates are received as an unbounded sequence of insertions and deletions of RDF triples, continuously maintaining query results. \\\hline

Querying Model and Expressiveness & Mainly, use of relational queries (e.g., C-SPARQL, CQELS) over sliding windows that limit the amount of data analyzed at any given time. & Event-based rule languages (e.g., EP-SPARQL) that support complex temporal conditions. &  BGP queries, which are continuously evaluated against a changing RDF graph by maintaining delta answers (incremental results computed over updates).\\\hline

Update Handling & \multicolumn{2}{|p{0.5\linewidth}|}{Processing new incoming data and not inherently handling deletions in the stream.}  &  Explicitly incorporates both insertions and deletions, maintaining delta embeddings to ensure query results are dynamically updated.\\\hline

State Management and Memory Usage & \multicolumn{2}{|p{0.5\linewidth}|}{Use of sliding windows to limit the amount of historical data kept in memory.}  & Aims to minimize cached data while still maintaining sufficient history to reconstruct query answers efficiently over an unbounded RDF graph. \\\hline

Pattern Matching/Query Execution & Continuous relational joins over streaming data in a window. & Detecting sequences of events with defined patterns and time constraints. & Graph pattern matching algorithms and specialized in-memory techniques to efficiently evaluate queries over evolving RDF graphs, ensuring real-time updates.  \\\hline
\end{tabular}
\caption{Comparison between the DSMS, CEP systems and the setting of this paper (Dynamic Linked Data Graphs - Dynamic LDG)}
\label{table:comparison-table}
\end{table}


In the aforementioned approaches, the streaming data mainly includes new triples; i.e., no deletion of graph triples is considered, as in the setting investigated in this paper. Capturing updates (including deletion delivered by the stream) of the streaming data has been investigated in the context of incremental reasoning (e.g., \cite{ren2011optimising,dell2014ch,barbieri2010incremental,volz2005incrementally,volz2005incrementally}). Incremental Materialization for RDF STreams (IMaRS) \cite{dell2014ch,barbieri2010incremental} considers streaming input annotated with expiration time, and uses the processing approach of C-SPARQL. Ren et al. in \cite{ren2011optimising,ren2010towards}, on the other hand, focus on more complex ontology languages and do not consider fixed time windows to estimate the expiration time. In context, it's worth noting the approach presented in~\cite{zhang2022handling}, where the authors propose an algorithm, called $IncTree_{RDF}$, to continuously compute the answers of a registered query over the streaming RDF graph. Although the framework does not typically support the deletion of RDF triples, the authors consider the expiration time for each edge in order to support a time-based sliding window model.

Continuous subgraph matching has also been investigated for non-RDF streaming graphs~\cite{wang2023survey}. The concept of expiration time is adopted in \cite{choi2018efficient} for computing subgraph matching for sliding windows. The authors also use a query decomposition approach based on node degree in graph streams. Fan et al. in \cite{fan2013incremental} investigated the incremental subgraph matching by finding a set of differences between the original matches and the new matches. In \cite{CHC+15}, the authors used a tree structure where the root node contains the query graph and the other nodes include subqueries of it. The leaf nodes of this structure include the edges of the query, and the structure is used to incrementally apply the streaming changes (partial matches are also maintained). In the same context, Graphflow \cite{kankanamge2017graphflow} and TurboFlux \cite{kim2018turboflux} present two approaches for the incremental computation of delta matches. Graphflow \cite{kankanamge2017graphflow} is based on a worst-case optimal join algorithm to evaluate the matching for each update, while TurboFlux \cite{kim2018turboflux} extends the input graph with additional edges, also taking into account the form of the query graph. This structure is used to find the delta matches efficiently.

As described in the related work, the vast majority of the works in the literature investigate a different setting (i.e., evaluating queries over streaming windows, e.g., C-SPARQL, EP-SPARQL, CQELS, IMaRS). Our setting/problem is more comparable with approaches such as Graphflow and TurboFlux, which, however, investigate different types of graphs and patterns. It’s worth noting that, among the related works over RDF streams, IMaRS~ \cite{dell2014ch,barbieri2010incremental} and $IncTree_{RDF}$~\cite{zhang2022handling} are the closest to our approach since they similarly handle deletions as well. However, IMaRS mainly focuses on maintaining the ontological entailments in the streaming window rather than minimizing the memory footprint and continuously updating the query answers based on the unbounded streaming graph. $IncTree_{RDF}$ focuses on computing the result over a sliding window by using a tree-like data structure, called $TStore$, to maintain the intermediate results and by using a query rewriting algorithm to rewrite the original query to a tree query.

In addition, the majority of related work focuses on providing expressive frameworks for querying streaming data (mainly using windows). The purpose and perspective of this work, however, differ in the sense that it aims to sacrifice expressiveness in order to ensure real-time response over the unbounded streaming graph (motivated examples include fraud detection and alarm mechanisms). In this context, we plan to extend our approach by also allowing aggregations.

\section{Preliminaries}
\label{sec:prelim}
	
In this section, we present the basic concepts used throughout this work.

\subsection{Data and query graphs}

Initially, we define two types of directed, labeled graphs with labeled edges that represent RDF data and Basic Graph Patterns (i.e., a subclass of SPARQL) over the RDF data~\cite{KGDN23-IS}. In the following, we consider two disjoint infinite sets $U_{so}$ and $U_p$ of URI references, an infinite set of (plain) literals
$L$ and a set of variables $V$.
	
A \emph{data triple}
is a triple of the form $(s, p, o)$, where $s$ take values from $U_{so}$, $o$ takes values from $(U_{so} \cup L)$ and $p$ takes values from $U_{p}$; i.e., $(s, p, o) \in  U_{so} \times U_p \times (U_{so} \cup L)$.
In each triple of this form, we say that $s$ is the \emph{subject}, $p$ is the \emph{predicate} and $o$ is the \emph{object} of $t$.
A \emph{data graph} $G$ is defined as a
set of data triples.
Similarly, we define a \emph{query triple pattern}\footnote{In this paper, we do not consider predicate variables.} (\emph{query triple} for short) as a triple $(s, p, o)$ in $ (U_{so} \cup V) \times U_p \times (U_{so} \cup L \cup V)$; i.e., $s$ could be either URI or variables from $V$ and $o$ could be either URI or variable, or literal.
%
A \emph{Basic Graph Pattern (BGP)}, or simply a \emph{query (graph)},  $Q$ is defined as a non-empty set of query triples.
In essence, each triple $(s, p, o)$ in the query and data graphs represents a directed edge from $s$ to $o$ which is labeled by $p$, while $s$ and $o$ represent nodes in the corresponding graphs.
The \emph{output pattern} $O(Q)$ of a query graph $Q$ is the tuple $(X_1, \dots, X_n)$, with $n \geq 0$, of all the variables appearing in $Q$ w.r.t. to a total order over the variables of $Q$
\footnote{Although there are $n!$ output patterns for each query of $n$ variables, we assume a predefined ordering given as part of query definition.}.
A query $Q$ is said to be a \emph{Boolean, or ground, query} if $n=0$ (i.e., there is not any variable in $Q$). The set of nodes of a data graph $G$ (resp. query graph $Q$) is denoted as $\CN(G)$ (resp.  $\CN(Q)$).
The set of variables of $Q$ is denoted as $\CV(Q)$.
In the following, we refer to either a URI or literal node as \textit{constant node}.
%

\subsection{Data and query graph decomposition}
	
In this subsection we define the notion of data and query graph decomposition.
	
\begin{definition}
\label{def:graphpartition}\label{def:graphdecomposition}
A \emph{data (resp. query) graph decomposition} ${\cal D}(F)$ of a data (resp. query) graph $F$
is an $m$-tuple of data (resp.  query) graphs ${\cal D}(F)=(F_1, \dots, F_m)$, with $m \geq 1$,  such that:
		
\begin{enumerate}
	\item $F_i \subseteq F$, for $i = 1, \dots, m$, and
	\item $\bigcup_i F_i = F$.
\end{enumerate}
		
Each data (resp. query) graph $F_i$ in a data (resp. query) graph decomposition is called a \emph{data (resp. query) graph segment}.
When, in a data/query graph decomposition, for all pairs $i$, $j$, with $1 \leq i < j \leq m$, it also holds $F_i \cap F_j = \emptyset$,
i.e. data (resp. query) graph segments are disjoint of each other,  then the data (resp. query) graph decomposition is said to be \emph{non-redundant} and the graph (resp. query) segments obtained form a partition of the triples of data (resp. query) graph $F$,
called \emph{$m$-triple partition} of  $F$.
\end{definition}

\subsection{Embeddings and query answers}

This subsection focuses on describing the main concepts used for the query evaluation. To compute the answers of a query $Q$ when it is posed on a data graph G, we consider finding proper mappings from the nodes and edges of $Q$ to the nodes and edges of $G$. Such kind of mappings are described by the concept of embedding, which is formally defined as follows.
	
\begin{definition}
\label{def:totalEmbedding}
A \emph{(total) embedding} of a query graph $Q$ in a data graph $G$ is a total mapping $e: {\cal{N}}(Q) \rightarrow {\cal{N}}(G)$ with the following properties:
\begin{enumerate}
	\item
		For each node $v \in  {\cal{N}}(Q)$, if $v$ is not a variable then $v = e(v)$.
	\item
		For each triple $(s, p, o) \in Q$, the triple  $(e(s), p, e(o))$ is in $G$.
\end{enumerate}
		
The tuple $(e(X_1), \dots, e(X_n))$, where $(X_1,\dots,X_n)$ is the output pattern
of $Q$, is said to be an \emph{answer}\footnote{The notion of \emph{answer} in this paper coincides with the term \emph{solution} used in SPARQL.} to the query $Q$. Notice that $e(Q) \subseteq G$.
The set containing the answers of a query $Q$ over a graph $G$ is denoted as $Q(G)$.
\end{definition}

	Note that the variables mapping \cite{perez2008semantics,perez2009semantics} considered for SPARQL evaluation is related with the concept of the embedding as follows. If $Q$ is a query pattern and $G$ is a data graph, and there is an embedding $e$ from ${\cal N}(Q)$ to ${\cal N}(G)$, then the mapping $\mu$ from ${\cal V}(Q)$ to ${\cal N}(G)$, so that $e(n)=\mu(n)$ for each variable node $n$ in ${\cal V}(Q)$, is a variable mapping.

\begin{definition}
\label{def:partialEmbedding}
A \emph{partial embedding} of a query graph $Q$ in a data graph $G$ is a total embedding of a subquery $Q'$ of $Q$ in $G$.
%
\end{definition}

In essence, a partial embedding represents a mapping from a subset of nodes and edges of  a query $Q$
to a given data graph $G$. In other words, partial embeddings represent partial answers to $Q$, provided that,
they can be appropriately ``combined" with other ``compatible" partial embeddings to give ``complete answers'' (i.e. total embeddings)  to the query $Q$.

\begin{definition}
\label{def:compatible}
Two partial embeddings $e_1: Q_1 \rightarrow G_1$ and $e_2: Q_2 \rightarrow G_2$ are said to be \emph{compatible} if
for every node $v \in  {\cal{N}}(Q_1) \cap {\cal{N}}(Q_2)$ such that $e_1(v)$ and $e_2(v)$ are defined, it holds that $e_1(v)= e_2(v)$.
\end{definition}
	
\begin{definition}
\label{def:join}
Let $e_1: Q_1 \rightarrow G_1$ and $e_2: Q_2 \rightarrow G_2$ be two compatible partial embeddings.
The \emph{join} of $e_1$ and $e_2$, denoted as $e_1   \otimes  e_2$, is the partial embedding  $e: Q_1 \cup Q_2 \rightarrow G_1 \cup G_2$ defined as follows:
	\[ e(v) = \left\{ \begin{array}{ll}
		e_1(v) & \mbox{if $e_1(v)$ is defined} \\
		e_2(v) & \mbox{if $e_2(v)$ is defined and $e_1(v)$ is undefined} \\
		\textup{undefined} & \mbox{if both $e_1(v)$ and $e_2(v)$ are undefined} \\
		\end{array}
	\right. \]
\end{definition}

\section{Special forms of queries and query decomposition}
\label{sec:special}

\subsection{Special forms of queries}

We now define two special classes of queries, the \emph{generalized star queries}
(\emph{star queries}, for short) and the \emph{var-connected queries}.

\begin{definition}
\label{def:srar-query-def}
A query $Q$ is called a \emph{generalized star query} if there exists a node $c \in {\cal N}(Q)$, called the \emph{central node} of $Q$
and denoted as $C(Q)$, such that for every triple $t = (s, p, o) \in Q$ it is either $s = c$ or $o = c$.
If the central node of $Q$ is a variable then $Q$ is called \emph{var-centric generalized star query}. A var-centric star query
is \emph{simple} if all the adjacent nodes to the central node are constants  (i.e., either URIs or literals).
\end{definition}
	
To define the class of var-connected queries we first define the notion of \emph{generalized path} connecting two nodes of the query $Q$.

\begin{definition}
	Let $Q$ be a BGP query and $v_1$, $v_{k+1}$, with $k \geq 1$, be two nodes in $Q$.
	A \emph{generalized path} between  $v_1$ and $v_{k+1}$  of length $k$, is a sequence of triples $t_1, \dots, t_k$  such that:
		
	\begin{enumerate}
		\item
			there is a sequence of nodes $v_2$, $\dots$, $v_{k}$ in $Q$, where $v_1, v_2, \dots, v_k, v_{k+1}$ are disjoint, and
			a sequence of predicates $p_1, \dots, p_k$ not necessarily distinct, and
		\item
			for each $i$, with $1 \leq i \leq k$, either $t_i = (v_i, p_i, v_{i+1})$, or $t_i = (v_{i+1}, p_i, v_{i})$.
	\end{enumerate}
		
	The \emph{distance} $d(v, v')$ between two nodes $v$ and $v'$ in $Q$,
	is the length of (i.e. the number of triples in) the shortest generalized path between $v$ and $v'$.
\end{definition}

\begin{definition}
\label{def:var-connected-query}
Consider a BGP query $Q$ with multiple variables (i.e., $|{\cal V}(Q)| > 1$) such that for
each pair of variables $(X, Y )$ of $Q$ there is a
generalized path connecting $X$ and $Y$ which does not contain any constant.
In such a case, we say that $Q$ is a \emph{var-connected query}.
\end{definition}
	
Notice that a var-centric query which is not simple is var-connected query.

\subsection{Connected-variable and star decompositions of BGP queries}
	
In this section, we present a non-redundant decomposition of a BGP query $Q$, called  \emph{connected-variable partition} of $Q$,  into a set of queries whose form facilitates
their efficient evaluation over a dynamic Linked Data graphs. In addition, we recall the star decomposition used and presented in \cite{KGDN23-IS} and \cite{kalogeros2020document}, which is going to be used to develop an efficient algorithm for evaluating var-connected queries.
	
\begin{definition}
\label{def:connected-variable-partition}
	Let $Q$ be a BGP query and $V$ be the set of variables in $Q$.
	A partition ${\cal P}(V)$ of the variables in $V$,  is said to be a \emph{connected-variable partition} of $V$ if the following hold:
		
	\begin{enumerate}
		\item For each $P_i \in {\cal P}(V)$ and for every two disjoint variables $u, v \in P_i$, there is a generalized path between   $u$ and $v$ in $Q$ whose nodes are  variables belonging to $P_i$.
		\item For each pair $P_i,  P_j \in {\cal P}(V)$, with $i \neq j$, there is no pair of variables $u$, $v$ such that $u \in P_i$ and $v \in P_j$ and there is a triple in $Q$ whose subject is one of these variables and whose object is the other.
	\end{enumerate}
\end{definition}

\begin{definition}
\label{def:cvd}
Let $Q$ be a BGP query, $V$ be the set of variables in $Q$, and ${\cal P}(V)$ be the connected-variable partition  of $V$.
The  \emph{connected-variable decomposition} ${\cal D_{CV}}(Q)$ of $Q$ is a non-redundant decomposition of $Q$ containing  $|{\cal P}(V)|$ non-ground queries (i.e. queries containing variables),  and (possibly) a ground query $Q_G$. These queries are constructed as follows:
		
	\begin{enumerate}
		\item
			For each element $P_i \in {\cal P}(V)$, we construct a query $Q_{P_i}$ = $\{t | t = (s, p, o) \in Q$ and either $s \in P_i$
			or $o \in P_i\}$.
		\item
			$Q_G =  Q - \bigcup_i Q_{P_i}$.
	\end{enumerate}
\end{definition}

Note that $Q_G$ may be empty. In this case $Q_G$ is not included in ${\cal D_{CV}}(Q)$.
Notice also that when, for some $i$, it holds that $|P_i| = 1$, then $Q_{P_i}$  is a simple var-centric star query.

\begin{example}
Consider the query $Q$ appearing in Fig.~\ref{fig:connected}. The set of variables of $Q$ is $V = \{?X, ?Y, ?Z, ?W\}$,
while the connected-variable partition ${\cal P}(V)$ of $V$ is ${\cal P}(V) = \{\{?X, ?Y, ?Z\}, \{?W\}\}$.
The connected-variable decomposition ${\cal D_{CV}}(Q)$ of $Q$ is ${\cal D_{CV}}(Q) = \{Q_1, Q_2, Q_G\}$. $Q_1$, $Q_2$ and $Q_G$ also appear in Fig.~\ref{fig:connected}.
		
\begin{figure*}[t]
	\centering
			\includegraphics[width=0.90\textwidth]{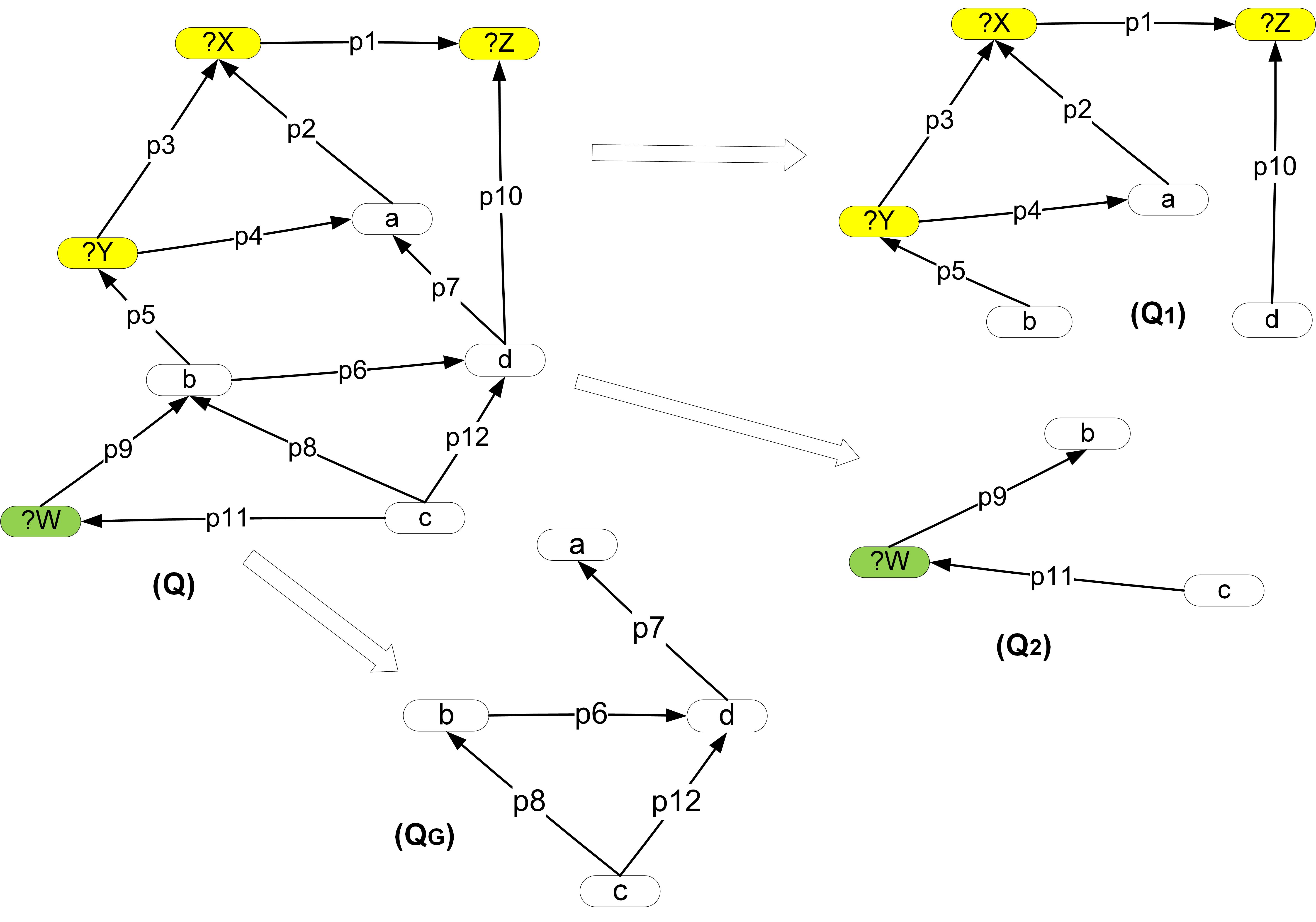}
	\caption{Connected-variable decomposition of a BGP query $Q$.}
	\label{fig:connected}
\end{figure*}
\end{example}

It should be noted that, when all members of ${\cal P}(V)$ are singletons,
all queries in ${\cal D_{CV}}(Q)$ containing variables are simple var-centric star queries.

\begin{definition}
\label{def:loosely-connected}
A BGP query $Q$ is said to be \emph{loosely-connected} if for each pair of disjoint variables $X$, $Y$ in $Q$, it holds that
every generalized path between $X$ and $Y$ contains at least one non-variable  node.
\end{definition}
	
\begin{lemma}
\label{lem:loosely_connected}
Let $Q$ be a loosely-connected BGP query and ${\cal D_{CV}}(Q)$ be the connected-variable decomposition of $Q$.
Then each non-ground query in ${\cal D_{CV}}(Q)$ is a simple var-centric star query.
\end{lemma}
	
\proof 
From Definitions~\ref{def:connected-variable-partition} and~\ref{def:loosely-connected}, we conclude that
each member of the connected-variable partition ${\cal P}(V)$ of the variables in ${\cal V}(Q)$
is  singleton.
The queries in ${\cal D_{CV}}(Q)$ are obtained by applying Definition~\ref{def:cvd}, thus, by construction, each query in ${\cal D_{CV}}(Q)$ has a variable as central node which is either the subject or the object of triples which have a non-variable object or subject, respectively. From Definition~\ref{def:srar-query-def}, we see that these queries are
simple var-centric star queries.
\QED
	
\begin{example} Consider the BGP query $Q$ appearing in Fig.~\ref{fig:loosely-connected}.
The set of variables of $Q$ is $V = \{?X, ?Y, ?Z\}$,
while the connected-variable partition ${\cal P}(V)$ of $V$ is ${\cal P}(V) = \{\{?X\}, \{?Y\}, \{?Z\}\}$.
The connected-variable decomposition ${\cal D_{CV}}(Q)$ of $Q$ is ${\cal D_{CV}}(Q) = \{Q_1$, $Q_2$, $Q_3$, $Q_G\}$,
where $Q_1$, $Q_2$, $Q_3$, and $Q_G$ appear in Fig.~\ref{fig:loosely-connected}.
We can see that  each query containing variables (i.e. $Q_1$, $Q_2$, and $Q_3$)
is a simple var-centric star query.

\begin{figure*}[t]
\centering
\includegraphics[width=0.90\textwidth]{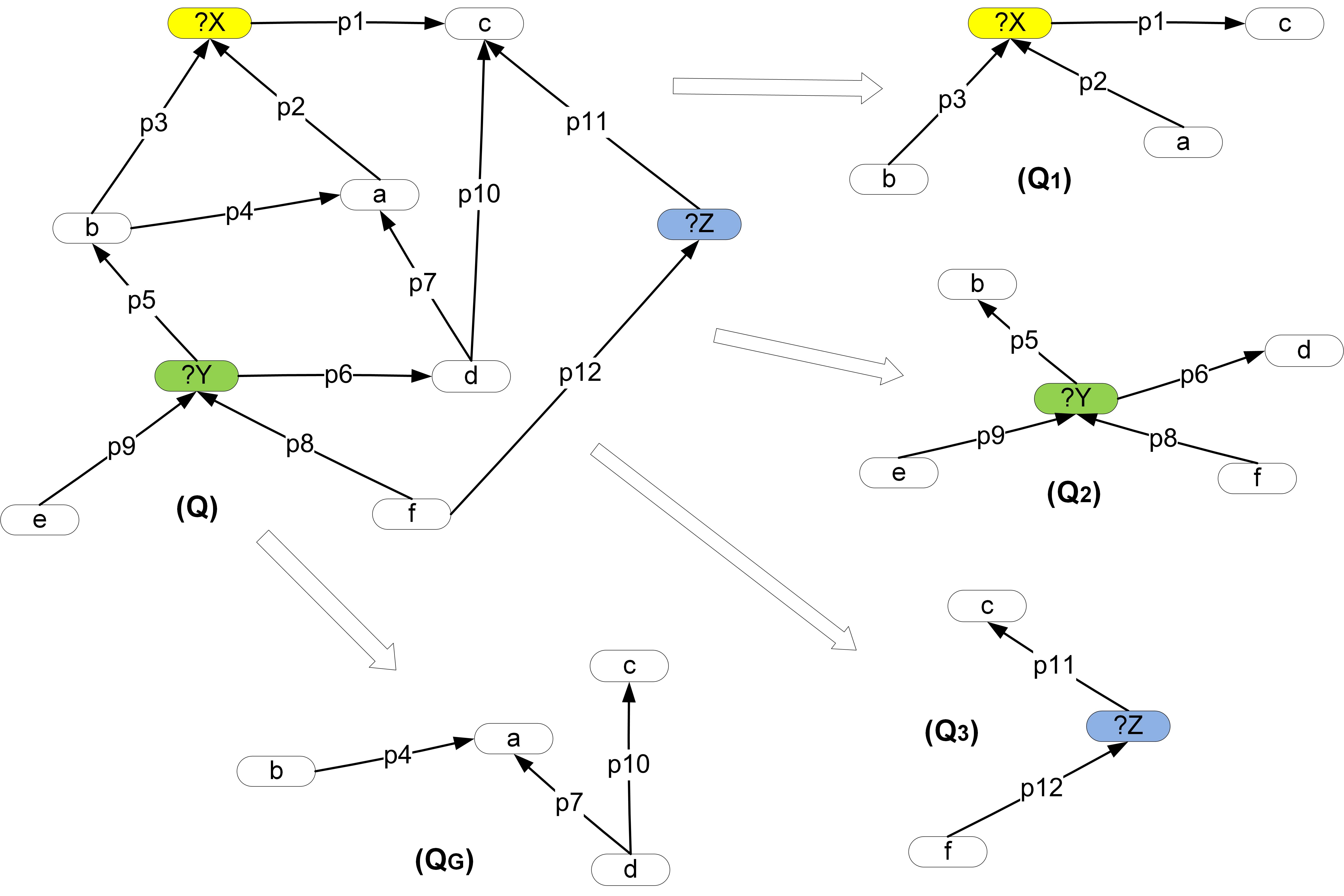}
\caption{Connected-variable decomposition of a loosely connected BGP query $Q$.}
\label{fig:loosely-connected}
\end{figure*}
\end{example}

Definition~\ref{def:graphdecomposition}   and Lemma~\ref{lem:loosely_connected} formally present the conditions under which a BGP query can be decomposed into a set of var-centric queries which, as we see in Section~\ref{sec:answer-queries}, can be efficiently evaluated. Lemma~\ref{lem:loosely_connected} shows that such a decomposition exists for any loosely-connected query.

\begin{definition}
\label{def:connected-variable-query}
Let $Q$ be a BGP query and $V$ be the set of variables in $Q$.
Assume that $|V| \geq 2$ and  ${\cal P}(V)$ is the   connected-variable partition  of $V$.
Then, if ${\cal P}(V)$ is a singleton, we say that $Q$ is a \emph{var-connected query}.
\end{definition}

Let us now see a different type of decomposition of a BGP query $Q$. In particular, the following definition presents the star decomposition which is also defined and used in \cite{KGDN23-IS} and \cite{kalogeros2020document}, and decomposes $Q$ into a set of generalized star queries with overlapping edges.
	
\begin{definition}\label{def:stardecomposition}
A \emph{star decomposition} of a query graph $Q$ is a decomposition ${\cal D}_Q = (Q_1, \dots, Q_m)$ of $Q$, with $m \geq 1$, s.t., if ${\cal N_P}=(N_1, \dots, N_m)$ is a partition of the nodes in ${\cal N}(Q)- C$, where $C$ is a set of constants, and
for each $i$, with $1 \leq i \leq m$, $Q_i = \{t \mid  t=(s, p, o)\ and\ t \in Q\ and\ s \in N_i\ or\ o \in N_i\}$.
A query triple $t = (s, p, o)\in Q_i$ is called \emph{replicated} if there is a subquery $Q_j\in\CD_Q$ s.t. $i\neq j$ and  $t\in Q_j$.
\end{definition}
	
As proved in \cite{kalogeros2020document} (Proposition 5.2) each BGP query $Q$ can be decomposed into a set of generalized star subqueries. Furthermore, joining the total embeddings of the subqueries in a star decomposition ${\cal D}_Q$ over an arbitrary data graph, we can find a total embedding of the query $Q$ (implied by Theorem 8 in \cite{KGDN23-IS}).
	
\begin{theorem}\label{th:QandGdecompandeval}
Let ${\cal D}_Q = (Q_1, \dots, Q_n)$ be a star decomposition of a BGP query $Q$.
Then, $e$ is a total embedding of $Q$ over a data graph $G$ if and only if $e$ is the join of $e_1, \dots, e_n$, where $e_1, \dots, e_n$ are mutually compatible embeddings such that for each $i$, with $1 \leq i \leq n$, $e_i$ is a total embedding of $Q_i$ over $G$.
\end{theorem}

\section{Continuous  pattern matching}
\label{sec:continuous}
	
In this work we consider data graphs
continuously changing (a.k.a, \textit{dynamic data graphs}), through infinite sequences of updates over the data.
In particular, we
consider an infinite, ordered sequence $\CM$ $=$ $(\DO_1$, $\DO_2$, $\dots)$ of update messages, called \emph{update stream}, of the form $\DO_k=<t_k, Op, e>$, where
\begin{itemize}
	\item $t_k$ is the time the message received, where  $k=1,2,\dots$,
	\item $Op$ is an update operation applied over a data graph and takes values from the following domain: $\{ins$, $del\}$, and
	\item $e=(s, p, o)$ is the edge to be either inserted or deleted (according to the operation specified by $Op$).
\end{itemize}
	
Note that the $ins$ operation stands for $insertion$, while $del$ stands for $deletion$. For example, the message $<t, ins, (s, p, o)>$, which is received at time $t$ describes an insertion of the edge $(s, p, o)$ into the data graph. The data graph resulted by applying an update message $\DO_k$ in $\CM$ over a data graph $G$ is a data graph $\DO_k(G)$ defined as follows:
$$\DO_k(G)= \left\{
\begin{array}{ll}
	G\cup\{(s, p, o))\}, & \quad O=ins \\
	G-\{(s, p, o)\}, & \quad O=del
\end{array}
\right .$$

Let $G_0$ be a data graph  at a certain time $t_0$.  At time $t_1$, we receive an update message $\DO_1$ in $\CM$. Applying the update $\DO_1$ on $G_0$, we get an updated graph $G_1$. Similarly, once we receive the update message $\DO_2$ in $\CM$ and apply it into $G_1$, we get the graph $G_2$.
Continuously applying all updates that are being received through $\CM$, we have a data graph that is slightly changing over time. We refer to such a graph $G=(G_0,G_1,\dots,G_k,\dots)$ as \textit{dynamic graph}, and to each data graph at a certain time as \textit{snapshot} of $G$. For simplicity, we denote the snapshot of $G$ at time $t_k$ as $G(t_k)$.
Hence, at time $t_{k}$, the graph snapshot $G_k$ of $G$ is given as follows:
$$\DO_{[0..k]}(G_{0}) = G_k = \DO_k(G_{k-1}) =\DO_k(\DO_{k-1}(\dots \DO_1(G_0)\dots )) $$

Considering a BGP query $Q$ which is continuously applied on each snapshot of the dynamic graph $G=(G_0,G_1,\dots,G_k,\dots)$, we might find different results at each time the query is evaluated. In particular, if $G_{k-1}$ is the graph snapshot at time $t_{k-1}$ and we receive an insertion message at time $t_k$, $Q(G_k)$ might contain embeddings that were not included in $Q(G_{k-1})$. In such a case (i.e., if $Q(G_k)\supset Q(G_{k-1})$), we say that each embedding in  $Q(G_k) - Q(G_{k-1})$ is a \textit{positive embedding}. Similarly, if we consider that the message received at time $t_k$ is a deletion, we might have embeddings in $Q(G_{k-1})$ that are no longer valid in $Q(G_k)$ (i.e., $Q(G_k)\subset Q(G_{k-1})$). If so, each embedding in $Q(G_{k-1}) - Q(G_k)$ is called \textit{negative embedding}. For simplicity, we refer to the set of positive and negative embeddings at time $t_k$ as \textit{delta embeddings}. In the following, we denote the sets of positive, negative and delta embeddings, at time $t_k$, as $Q^+(G_k)$, $Q^-(G_k)$ and  $Q^{\delta}(G_k)$, respectively.
	
\begin{definition} \textbf{(Problem Definition)}
Considering a dynamic graph $G=(G_0,G_1,\dots,G_k,\dots)$, an update stream $\CM = (\DO_1,\DO_2,\dots)$, and a query  $Q$,
we want to find for each time $t_k$, with $k=1,2,\dots$, the outputs of all the delta embeddings  $Q^{\delta}(G_k) = Q^+(G_k)\cup Q^-(G_k)$, where
\begin{itemize}
	\item $G_k=\DO_{[0..k]}(G_{0})$,
	\item the positive embeddings is given as $Q^+(G_k)=Q(G_k) - Q(G_{k-1})$,
	\item the negative embeddings is given as $Q^-(G_k)=Q(G_{k-1}) - Q(G_k)$.
\end{itemize}
\end{definition}

The following sections are based on the following assumptions: (a) Whenever  an update message of the form $<t_k, del, e>$ is received, we assume that $e \in  G_{k-1}$, and (b) Whenever  an update message of the form $<t_k, ins, e>$ is received, we assume that $e \notin G_{k-1}$.

\section{Answering BGP queries over dynamic Linked Data}
\label{sec:answer-queries}
	
In this section, we investigate methods for continuously answering BGP queries over dynamic graphs. In particular, we focus on finding the delta embeddings for each message received from an update stream. The set of embeddings found by each of the algorithms presented in the subsequent sections ensure that collecting all the delta embeddings from the beginning of the stream to the time $t$ and applying the corresponding operations (deletions and insertions) according to the order they found, the remaining embeddings describe the answer of the given query over the graph $G_t$.
	Initially, we focus on certain well-used subclasses of BGP queries, defined in the previous sections,  which include ground BGP queries, simple var-centric star queries, and loosely-connected BGP queries, and present efficient algorithms for computing delta embeddings and investigate time and space complexity.
	Then, we move forward and present an algorithm aiming to efficiently solve the problem for any BGP query by initially handling the var-connected queries.
	
Although, at a certain time, to find all the embeddings of a BGP query takes exponential time in the query size, since the query is predefined and does not change over time, the answers can be computed in polynomial time. For certain subclasses (e.g., ground and simple var centric queries), we can improve the complexity further by leveraging structural properties of the query and optimizing the evaluation process.

\subsection{A generic query evaluation procedure}
\label{subsec:generic}

	Starting our analysis, we initially focus on the intuition of the evaluation algorithms presented in the subsequent section. The methodology used to construct the algorithms relies on the proper decomposition of the given query into a set of subqueries. The following lemma shows that finding a connected-variable decomposition of a query, the total embeddings of the subqueries give a total embedding of the initial query.  	
	
\begin{lemma} \label{lem:eval1}
Let $G$ be a data graph and  $Q$ be a BGP query. Assume that ${\cal D_{CV}}(Q)$ = $\{Q_1, Q_2, \dots, Q_n\}$
is the connected-variable decomposition of $Q$.
Assume also that $e_1, e_2, \dots, e_n$ are total embeddings of $Q_1, Q_2, \dots, Q_n$ respectively, in $G$.
Then, $e_1, e_2, \dots, e_n$ are compatible partial embeddings of $Q$ in $G$ and
$e =  e_1 \otimes e_2 \otimes \dots \otimes e_n$ is a total embedding of $Q$ in $G$.
\end{lemma}

\proof
From  Definitions~\ref{def:totalEmbedding} and~\ref{def:partialEmbedding}  we see that a total embedding of a subquery $Q_i$ of $Q$ is also a partial embedding of $Q$.
Definition~\ref{def:cvd} implies that $V(Q_i) \cap V(Q_j)= \emptyset$
for each pair of queries $Q_i$ and $Q_j$, with $i \neq j$, in ${\cal D_{CV}}(Q)$.
Thus, as queries $Q_1, \dots, Q_n$ share no common variables, from Definition~\ref{def:compatible} we conclude that
$e_1, e_2, \dots, e_n$ are compatible partial embeddings of $Q$.
By joining $e_1, e_2, \dots, e_n$ (as described in Definition~\ref{def:join}) we get the  embedding $e =  e_1 \otimes e_2 \otimes \dots \otimes e_n$ of $Q$ in $G$ which is total as each node and edge  of $Q$ is covered by some $e_i$ as  $e_1, e_2, \dots, e_n$ are total embeddings of the corresponding subqueries in  ${\cal D_{CV}}(Q)$ and, as we can easily derive from Definition~\ref{def:cvd}, ${\cal D_{CV}}(Q)$ covers all nodes and edges of $Q$.
\QED

\begin{lemma} \label{lem:eval2}
Let $G$ be a data graph and  $Q$ be a BGP query and ${\cal D_{CV}}(Q)$ = $\{Q_1, Q_2, \dots, Q_n\}$
be the connected-variable decomposition of $Q$. Assume that $e$ is a total embedding of $Q$ in $G$. Then, there are total embeddings
$e_1, e_2, \dots, e_n$ of $Q_1, Q_2, \dots, Q_n$ respectively, in $G$ such that
$e =  e_1 \otimes e_2 \otimes \dots \otimes e_n$.
\end{lemma}

\proof
For each $Q_i$, with $1 \leq i \leq n$, we construct the embedding $e_i$ obtained by restricting $e$ to the mappings of the nodes and edges in $Q_i$. As, by construction, $Q_i \subseteq Q$ and $e$ is a total embedding of $Q$ on $G$, we conclude that $e_i$ is also a total embedding of $Q_i$ to $G$.
By construction $e_1, e_2, \dots, e_n$ are compatible partial embeddings of $Q$ in $G$.
Besides, Definition~\ref{def:cvd} implies that $Q_1 \cup Q_2 \cup \dots \cup Q_n = Q$. Therefore, $e =  e_1 \otimes e_2 \otimes \dots \otimes e_n$.
\QED

	Based on the Lemmas~\ref{lem:eval1} and~\ref{lem:eval2}, the following generic procedure  can be used for the evaluation of every BGP query $Q$:
	
\begin{enumerate}
\item 
$Q$ is decomposed by applying Definition~\ref{def:cvd} to obtain its connected-variable decomposition ${\cal D_{CV}}(Q)$.
\item 
All subqueries in ${\cal D_{CV}}(Q)$ are evaluated independently of each other.
\item 
The set of answers of the query $Q$ is the Cartesian product of the sets of answers of the subqueries in ${\cal D_{CV}}(Q)$.
\end{enumerate}

	Concerning the second step of the above procedure, it is important to observe that ${\cal D_{CV}}(Q)$ may contain the following three different types of queries (subqueries of the query $Q$) and therefore it is necessary to design algorithms for the  evaluation of each query type:
	
\begin{enumerate}
\item [(a)] zero or one ground BGP query.
\item [(b)] a (possibly empty) set of simple var-centric star queries, and
\item [(c)] a (possibly empty) set of var-connected queries.
Notice that, in this type of queries, every query triple has at least one variable (as subject or object).
\end{enumerate}
	
In the following section, we propose algorithms to evaluate queries belonging to each one of the above query types.
However, as an algorithm for the evaluation of queries of type (c) is complex, we present an algorithm that evaluates an intermediate type of queries, more specifically, the loosely connected queries.
	
\subsection{Evaluating ground BGP queries}
\label{subsec:ground}

A BGP query $Q_G$  is true in the current time $t_c$
whenever for each triple in $Q_G$ an insert message has received and this inserted triple has not been deleted by a subsequent delete message.
To keep track of the triples of $Q_G$ that are true at time $t_c$, we employ an array $M_{Q_G}$ of $|Q_G|$ boolean variables,
called \emph{triple match state} of $Q_G$. If $M_{Q_G}$ has the value $1$ at each position, i.e., $M_{Q_G}[i]=1$ for all $i$, then we say that $M_{Q_G}$ is \textit{complete}.
We also assume an enumeration $e_1, e_2, \dots, e_{|Q_G|}$ of the triples in $Q_G$.
$M_{Q_G}[i] = 1$ if the triple $e_i$ of $Q_G$ appears in the current state of $Q_G$;
(i.e. an insert message for $e_i$ has arrived and $e_i$ has not be deleted by a subsequent delete message);
otherwise, $M_{Q_G}[i] = 0$. Our algorithm is illustrated in Algorithm~\ref{alg:ground}. Note that, auxiliary functions used in algorithms~\ref{alg:ground},  \ref{alg:var-centric} and \ref{alg:loosely}, are collectively presented in Algorithm~\ref{alg:functions}.

	The space and time complexity of Algorithm~\ref{alg:ground} is given in Proposition~\ref{prop:ground-complexity}.
	
 \begin{proposition}
\label{prop:ground-complexity}
Considering a ground BGP query $Q_G$, Algorithm~\ref{alg:ground} requires $O(|Q_G|)$ space and it computes the delta embed\-dings for each input message in $O(1)$ time. The complexities are given in terms of the query graph (which is the input of the evaluation algorithm).
\end{proposition}

\proof
Let $Q_G$ be a ground query and $\CU_k=<t_k,Op,e>$ an update message received
at  time $t_k$.
The space required by the Algorithm~\ref{alg:ground}, is required for storing/caching the array $M_{Q_G}$. By construction of $M_{Q_G}$ (line 2), it requires $|M_{Q_G}|=|Q_G|$ bits, which is constant for the given $Q_G$. Hence, it requires $O(|Q_G|)=O(1)$ space.

To find the delta embeddings for $Q_G$, we search all the triples of $Q_G$ (lines 9, 14).
Since the number of edges of the query is fixed, due to the index over $M_{Q_G}$ (i.e., the $i^{th}$ triple in $Q_G$ requires an update over the $i^{th}$ bit of  $M_{Q_G}$), Algorithm~\ref{alg:ground} requires constant time to compute the answers; hence,
the computation of the delta embedding is performed in constant time; i.e., $O(1)$ time.
\QED

\subsection{Evaluating simple var-centric star queries}
\label{subsec:var-centric}
	
The procedure for the evaluation of a simple var-centric star query over an input stream is based on the following observation:
the central node of a simple var-centric star query $Q_S$ is a variable $X$ common to all triples in $Q_S$
(either as the subject or as the object of the triple), while all the other nodes in $Q_S$  are either URI's or literals.
Hence, the first edge $e$ of $G$, received through an insertion message, that `matches' with a triple in $Q_S$, instantiates the variable $X$ to a constant value $X_G$.
In this way, we get a ground instance $Q_G$ of $Q_S$ whose evaluation can proceed in a similar way as in Algorithm~\ref{alg:ground} presented in Subsection~\ref{subsec:ground}.
For the presentation of the evaluation procedure, we consider a list $L$
of triples of the form $(X_G, Q_G, M_{Q_G})$, called \emph{list of ground var-centric instances} of $Q_S$, where $X_G$ is an instance of the variable $X$ and $Q_G$ is the ground instance of $Q_S$
obtained by replacing $X$ by  $X_G$, and $M_{Q_G}$ is the triple match state of $Q_G$.
When an insert update message results in a new instantiation of the variable $X$, a new ground instance of $Q_S$ is created and a new triple of the form
$(X_G, Q_G, M_{Q_G})$, is added to $L$. On the other hand, when  a delete update message results in the deletion of  the last true triple in a ground instance $Q'_G$ of $Q_S$ the corresponding triple $(X'_G, Q'_G, M_{Q'_G})$ is removed from $L$.
When, an insert update message, turns to true the last non true boolean variables in the triple match state of the corresponding query $Q_G$, the algorithm returns a positive answer  $<Q_S, positive, (X_G)>$.
On the other hand, if a delete update message turns to false a boolean variables in a triple match state of a query $Q_G$, where all boolean variables of $M_{Q_G}$ were true before the hose all boolean variables  of the corresponding query $Q_G$, which before
this delete message, then  the algorithm returns a negative answer $<Q_S, negative, (X_G)>$.
The complete algorithm is depicted by Algorithm~\ref{alg:var-centric}. Note that implementing the list of ground var-centric instances through a Hash Map we can significantly improve the query evaluation performance.
	
		{\scriptsize
		\begin{algorithm}[H]
			\caption{Algorithm that Evaluates ground BGP queries}
			\label{alg:ground}
			\SetAlgoLined
			\DontPrintSemicolon
			\SetKwFor{Loop}{Loop}{}{EndLoop}
			\SetKwInOut{Output}{output}
			
			\textbf{Procedure} eval\_groud\_BGP\_query($Q_G$)\;
			\PrintSemicolon
			\KwData{$Q_G$: a ground query;}
			\KwResult{Positive/negative answers;}
			
			\lFor{$i = 1$  \KwTo  $|Q_G|$} {
				$M_{Q_G}[i] = 0$
			}
			
			\Loop{}
			{Get the next update message ${\cal U}_k$\;
				R = $process\_update\_message\_on\_ground\_BGP$(${\cal U}_k$, $Q_G$, $M_{Q_G}$)\;
				\If{$R = <Q_G, positive,()>$ \textbf{or} $R = <Q_G, negative,()>$} {\Output{$R$} }
			}
			
			\BlankLine
			\DontPrintSemicolon
			
			\textbf{Procedure} process\_update\_message\_on\_ground\_BGP(${\cal U}_k$, $Q_G$, $M_{Q_G}$)\;
			\PrintSemicolon
			\KwData{${\cal U}_k$: an update message; $Q_G$: a ground query; $M_{Q_G}$: the triple match state of $Q_G$;}
			\KwResult{A positive/negative/no\_new\_embedding message;}
			\eIf{${\cal U}_k =  <k, ins, e>$}  {
				\If{$e$ = $e_i$, for some $e_i \in G_G$, with $1 \leq i \leq |Q_G|$} {
					$M_{Q_G}[i] = 1$\;
					\lIf{ground\_BGP\_eval$(M_{Q_G}, Q_G)$ = true} {\KwRet{$<Q_G, positive,()>$}}
				}
			}
			{
				\If{${\cal U}_k =  <k,  del, e>$} {
					\If{$e$ = $e_i$, for some $e_i \in G_G$, with $1 \leq i \leq |Q_G|$} {
						\eIf{ground\_BGP\_eval($M_{Q_G}$, $Q_G$)  = true } {
							$M_{Q_G}[i] = 0$ \;
							\KwRet{$<Q_G, negative,()>$}}
						{$M_{Q_G}[i] = 0$\; }
					}
				}
			}
			\KwRet{$<Q_G, no\_new\_embedding, ()>$} \;
		\end{algorithm}
	}

 \begin{proposition}
\label{prop:var-centric-complexity}
Considering a simple var-centric star query $Q_S$, Algorithm~\ref{alg:var-centric} requires $O(n)$ space, where $n$ is the number of the nodes of the data graph,  and it computes the delta embeddings for each input message in $O(2|Q_S|)+O(1)$ time.
\end{proposition}
\proof
Let $Q_S$ be a simple var-centric star query with central variable $X$ and $\CU_k$ = $<t_k, Op, e>$ an update message received
at time $t_k$. Also, assume that $G=(G_0,G_1,\dots,G_k)$ is the dynamic graph at time $t_k$.
Algorithm~\ref{alg:var-centric} requires space for storing the list $L$. To compute the size of $L$, we notice that $L$ contains a single triple of the form $(X_G, Q_G, M_{Q_G})$ for each different instance $X_G$ of the central node $X$ (line 18), where $M_{Q_G}$ is bit array of length $|Q_G|$ and $Q_G$ is the instance of $Q_S$ obtained by replacing $X$ by $X_G$. If $n$ equals the number of nodes of ${\cal{N}}(G_{k})$ (i.e., all the nodes of the edges that have been inserted and not been deleted from $t_0$ to $t_{k}$), then, in an extreme scenario, $X$ maps on each node in ${\cal{N}}(G_{k})$; i.e.,
the size of $L$ equals $n=|{\cal{N}}(G_{k})|$. Thus, the required space from Algorithm~\ref{alg:var-centric} is $O(n)$.

To find the delta embeddings for $Q_S$, we
initially search $Q_S$ to find if there is any triple that $e$ unifies with (lines 9, 22).
This takes $O(|Q_S|)$ time. If  $e$ unifies with a triple $e_j\in Q_S$ (either lines 9-19, or lines 22-28), we search the list $L$ for instances of $X$. Implementing $L$ as Hash Map using the instances of $X$ as keys, the complexity for searching the list is $O(1)$. If we find a triple $(X_G, Q_G, M_{Q_G})$ in  $L$, we also need $O(|Q_S|)$ time to update the $M_{Q_G}$ (lines 16-17, 26), and return each delta embeddings. Thus, the time of computing the delta embeddings is $O(2|Q_S|)+O(1)$.
\QED

	{\scriptsize
	\begin{algorithm}[H]
		\SetAlgoLined
		\SetKwFor{Loop}{Loop}{}{EndLoop}
		\SetKwInOut{Output}{output}
		
		\BlankLine
		\DontPrintSemicolon

		\textbf{Procedure}  eval\_simple\_var-centric\_star\_query($Q_S$)\;
		\PrintSemicolon
		\KwData{$Q_S$: a simple var-centric star query;}
		\KwResult{positive/negative answers;}
		
		$L = \{\}$\;
		
		\Loop{}
		{Get the next update message ${\cal U}_k$\;
			R = $process\_update\_message\_on\_star$(${\cal U}_k$, $Q_S$, $L$)\;
			\If{$R = <Q_S, positive,(X_Q)>$ \textbf{or} $R = <Q_S, negative,(X_Q)>$} {\Output{$R$} }
		}

		\BlankLine
		\DontPrintSemicolon

		\textbf{Procedure} process\_update\_message\_on\_star(${\cal U}_k$, $Q_S$, $L$)\;
		\PrintSemicolon
		\KwData{${\cal U}_k$: an update message; $Q_S$: a simple var-centric query; $L$: list of ground var-centric instances of $Q_S$;}
		\KwResult{A positive/negative/no\_new\_embedding message;}

		\eIf{${\cal U}_k =  <k, ins, e>$}  {
			\If{$e$ unifies with $e_j$ in $Q_S$, for some $j$, with $1 \leq j \leq |Q_S|$ } {
				let $X_G$ be the value of $X$ obtained by this unification\;
				let $Q_G$ be the instance of $Q_S$ obtained by replacing $X$ by $X_G$\;
				\eIf{there is a triple $(X_G, Q_G, M_{Q_G}) \in L$  for some  $M_{Q_G}$ }  {
					$M_{Q_G}[j] =  1$\;
					\lIf{ground\_BGP\_eval($M_{Q_G}$, $Q_G$) = true} {\KwRet{$<Q_S, positive,(X_G)>$}}
				}
				{
					\lFor{$i = 1$  \KwTo  $|Q_S|$} {$M_{Q_G}[i] = 0$}
					$M_{Q_G}[j] = 1$\;
					$L = L \cup \{(X_G, Q_G, M_{Q_G})\}$\;
					\lIf{$|Q_S| = 1$} {\KwRet{$<Q_S, positive, (X_G)>$ }}
				}
			}
		}
		{
			\If {${\cal U}_k =  <k, del, e>$} {
				\If{$e$ unifies with $e_i$ in $Q_S$, for some $i$, with $1 \leq i \leq |Q_S|$}  {
					\ForEach{$m = (X_G, Q_G, M_{Q_G}) \in L$} {
						\If{there is a triple $e_j \in Q_G$ such that $e = e_j$}  {
							\lIf{ground\_BGP\_eval($M_{Q_G}$, $Q_G$) = true}{\KwRet{$<Q_S, negative,(X_G)>$}}
							$M_{Q_G}[j] = 0$\;
							\ForEach{$m = (X'_G, Q'_G, M'_{Q'_G}) \in L$} {
								\lIf{all\_zero($M'_{Q'_G}, Q'_G$) = true} {remove $m$ from $L$}
							}
						}
					}
				}
			}
		}
		\KwRet{$<Q_S, no\_new\_embedding, ()>$}
		\caption{Algorithm that evaluates simple var-centric star queries}
		\label{alg:var-centric}
	\end{algorithm}
}
	
\subsection{Evaluating loosely-connected BGP queries}
\label{subsec:loosely-connected}

Let $Q$ be a loosely-connected BGP query and ${\cal D_{CV}}(Q)$ its connected-variable decomposition.
Lemma~\ref{lem:loosely_connected} implies that each non-ground query in ${\cal D_{CV}}(Q)$ is a simple var-centric star query.
To evaluate $Q$ it suffices to evaluate all queries in ${\cal D_{CV}}(Q)$
and then compute the Cartesian product of the answers.
Algorithm~\ref{alg:loosely} presents the continuous evaluation of such queries.
The algorithm caches for each star subquery all the values instantiating its central node. This minimizes the amount of data that is required in order to find any of the delta embeddings.
	

\begin{proposition}
\label{prop:loosely-connected-complexity}
Considering a loosely-connected query $Q_L$ and its connected-variable decomposition ${\cal D_{CV}}(Q_L)$, Algorithm~\ref{alg:loosely} needs $O(mn+|Q_G|)$ space, and $O(n+2m|Q_S|)$ time to compute delta embeddings of an input message,  where ${\cal D_{CV}}(Q_L)$ contains $m$ simple var-centric subqueries and the ground query $Q_G$, $n$ is the number of nodes of the data graph, and $Q_S$ is the simple var-centric subqueries in ${\cal D_{CV}}(Q_L)$ with the maximum number of nodes.
\end{proposition}

\proof
Let $Q_L$ be a loosely-connected query, $\CU_k=<t_k,Op,e>$ an update message received at time $t_k$, and $G=(G_0,G_1,\dots,G_k)$ is the dynamic graph at time $t_k$, and ${\cal D_{CV}}(Q_L)$ is the connected-variable decomposition of $Q_L$.
Algorithm~\ref{alg:loosely} uses Algorithm~\ref{alg:var-centric} (line 21) and Algorithm~\ref{alg:ground} (line 11) to compute the delta embeddings of each var-centric $Q_S\in {\cal D_{CV}}(Q_L)$ and the ground $Q_G\in {\cal D_{CV}}(Q_L)$, respectively. Hence, if
${\cal D_{CV}}(Q_L)$ contains $m+1$ subqueries (i.e., $m+1=|{\cal D_{CV}}(Q_L)|$), where one of them is a ground query, and $n = |{\cal{N}}(G_k)|$,
then Algorithm~\ref{alg:loosely} requires $(m-1)\cdot O(n)$ space to store the lists of the simple var-centric subqueries in ${\cal D_{CV}}(Q_L)$ (implied by Proposition~\ref{prop:var-centric-complexity}) and $O(|Q_G|)$ to store the single bit array required for computing the embeddings of the ground query $Q_G$ in ${\cal D_{CV}}(Q_L)$ (implied by Proposition~\ref{prop:ground-complexity}); i.e., it requires $m\cdot O(n)+O(|Q_G|)=O(mn)+O(|Q_G|)$ space. Algorithm~\ref{alg:loosely} also stores into the list $T_S$ the instances of the central variable of each var-centric query $Q_S$ given by each total embedding of $Q_S$ found till time $t_k$; the required space is included into $O(mn)$ space computed previously. Thus, the algorithm requires $O(mn)+O(|Q_G|) = O(mn+|Q_G|)$ space.

To compute the time required to find the delta embeddings and update the required structures (list of each var-centric subquery and the bit array for the ground subquery), we follow a similar approach. In particular, the time required is given by the time to compute the partial embeddings for var-centric and ground queries in ${\cal D_{CV}}(Q_L)$; which is $O(2m|Q_S|)+O(1)$, where $Q_S$ is the simple var-centric subquery in ${\cal D_{CV}}(Q_L)$ with the maximum number of nodes. In a worst-case scenario, the algorithm requires $O(n)$  to compute the Cartesian product of the partial embeddings.
Thus, it computes the delta embeddings for each input message in $O(n)+O(2m|Q_S|)+O(1)=O(n+2m|Q_S|)$ time.
\QED


	{\scriptsize	
		\begin{algorithm}[H]
			\SetAlgoLined
			\SetAlgoLined
			\SetKwFor{Loop}{Loop}{}{EndLoop}
			\SetKwInOut{Output}{output}
			
			\BlankLine
			\DontPrintSemicolon

			\textbf{Procedure}  eval\_loosely-connected\_query($Q_L$)\;
			\PrintSemicolon
			\KwData{$Q_L$: a loosely connected query;}
			\KwResult{positive/negative answers;}
			
			Let ${\cal D_{CV}}(Q_L)$ be the connected-variable decomposition of $Q_L$\;
			Let $<G,S>$ be a partition of ${\cal D_{CV}}(Q_L)$ s.t. $G$ contains the ground
			queries in ${\cal D_{CV}}(Q_L)$  and $S$ contains the simple var-centric queries in ${\cal D_{CV}}(Q_L)$\;
			
			\lFor{$i = 1$  \KwTo  $|G|$} {$T_{G}[i] = 0$}
			\For{$i = 1$  \KwTo  $|G|$} {
				\lFor{$j = 1$  \KwTo  $|G[i]|$} {$M_{G[i]}[j] = 0$}
			}
			\lFor{$i = 1$  \KwTo  $|S|$} {$T_{S}[i] = []$}

			\Loop{}
			{
				Get the next update message ${\cal U}_k$\;
				\For{$i = 1$  \KwTo  $|G|$} {
					R = $process\_update\_message\_on\_ground\_BGP$(${\cal U}_k$, $G[i]$, $M_{G[i]}$)\;
					\If{$R = < G[i], positive, ()>$} {
						$T_{G}[i] = 1$\;
						\If{$all\_ground\_true(G, T_G) = true$ \textbf{and} $all\_var-centric\_true(S, T_{S}) = true$
							(i.e. each var centric query in $S$ has at least one answer)} {
							\ForEach{tuple $O_S$ in the cartesian product of output patterns of the queries in $S$} {
								\Output{$<Q_L, positive, O_S>$}
							}
						}
					}
					\ElseIf{$R = < G[i], negative, ()>$} {
						\If{$all\_ground\_true(G, T_G) = true$ \textbf{and} $all\_var-centric\_true(S, T_{S}) = true$} {
							\ForEach{tuple $O_S$ in the cartesian product of output patterns of the queries in $S$} {
								\Output{$< Q_L, negative, O_S>$}
							}
						}
						$T_{G}[i] = 0$\;
					}
				}
				
				\For{$i = 1$  \KwTo  $|S|$ } {
					R = $process\_update\_message\_on\_star$(${\cal U}_k$, $S[i]$, $T_{S}[i]$)\;
					\If{$R = < S[i], positive, (X'_{S[i]})>$} {
						$T_{S}[i] = T_{S}[i] \cup \{X'_{S[i]}\}$\;
						\If{$all\_ground\_true(G, T_G) = true$ \textbf{and} $all\_var-centric\_true(S, T_{S}) = true$
							(i.e. each var centric query in $S$ has at least one answer)} {
							$temp = T_{S}[i]$\;
							$T_{S}[i] = \{X'_{S[i]}\}$\;
							\ForEach{tuple $V_t$ in the cartesian product  of the sets in $T_{S}$} {
								\Output{$< Q_L, positive, V_t >$}
							}
							
							$T_{S}[i]= temp$\;
						}
					}
					\ElseIf{$R = < S[i], negative, (X'_{S[i]})>$} {
						\If{$all\_ground\_true(G, T_G) = true$ \textbf{and} $all\_var-centric\_true(S, T_{S}) = true$ (i.e. all queries in $Q_L$ have an answer (before the receipt of the  message))|}  {
							$temp = T_{S}[i]$\;
							$T_S[i] = \{X'_{S[i]}\}$\;
							\ForEach{tuple $V_t$ in the cartesian product  of the sets in $T_{S}$} {
								\Output{$< Q_L, negative, V_t >$}
							}
							$T_S[i] = temp$\;
						}
						$T_S[i] = T_S[i] - \{X'_{S[i]}\}$\;
					}
				}
			}
		\caption{Algorithm that evaluates loosely-connected queries}
			\label{alg:loosely}
		\end{algorithm}
	}
	
	
	 $\;$
	
	{\scriptsize
		\begin{algorithm}[H]
			\SetAlgoLined
			\SetAlgoLined
			\SetKwFor{Loop}{Loop}{}{EndLoop}
			\SetKwInOut{Output}{output}
			
			\BlankLine
			\DontPrintSemicolon
			
			\textbf{Function} ground\_BGP\_eval($M_{Q}$, $Q$)\;
			\PrintSemicolon
			\KwData{$M_{Q}$: the triple match state of $Q$;}
			\For{$i = 1$  \KwTo  $|Q|$} {
				\lIf{$M_{Q}[i] = 0$} {\KwRet{$false$}}
			}
			\KwRet{$true$}\;

			\BlankLine
			\DontPrintSemicolon

			\textbf{Function}  all\_zero($M_{Q_G}$, $Q_G$)\;
			\PrintSemicolon
			\KwData{$M_{Q_G}$: the triple match state of the query $Q_G$;}
			\For{$i = 1$  \KwTo  $|Q_G|$} {
				\lIf{$M_{Q_G}[i] = 1$} {\KwRet{$false$}}
			}
			\KwRet{$true$}\;
			
			\BlankLine
			\DontPrintSemicolon

			\textbf{Function} $all\_ground\_true$($G$, $T_G$)\;
			\PrintSemicolon
			\KwData{$G$: a list of ground queries; $T_{G}$: array of flags showing which ground queries are true;}
			
			\For{$i = 1$  \KwTo  $|G|$} {
				\lIf{$T_{G}[i] = 0$} {\KwRet{$false$}}
			}
			\KwRet{$true$}\;
			
			\BlankLine
			\DontPrintSemicolon
			
			\textbf{Function} $all\_var-centric\_true$($S$, $T_S$)\;
			\PrintSemicolon
			\KwData{$S$: a list of simple var-centric queries; $T_{S}$: array of sets; each set containing the output values of var centric queries;}

			\For{$i = 1$  \KwTo  $|S|$}  {
				\lIf{$T_{S}[i] = []$} {\KwRet{$false$}}
			}
			\KwRet{$true$}\;
			\caption{Functions used in algorithms~\ref{alg:ground}, \ref{alg:var-centric} and \ref{alg:loosely}}
			\label{alg:functions}
		\end{algorithm}
	}
	



\subsection{Evaluating var-connected queries}
\label{sec:eval-connected-var-queries}
\begin{figure}
		\begin{center}
			\includegraphics[width=0.9\linewidth]{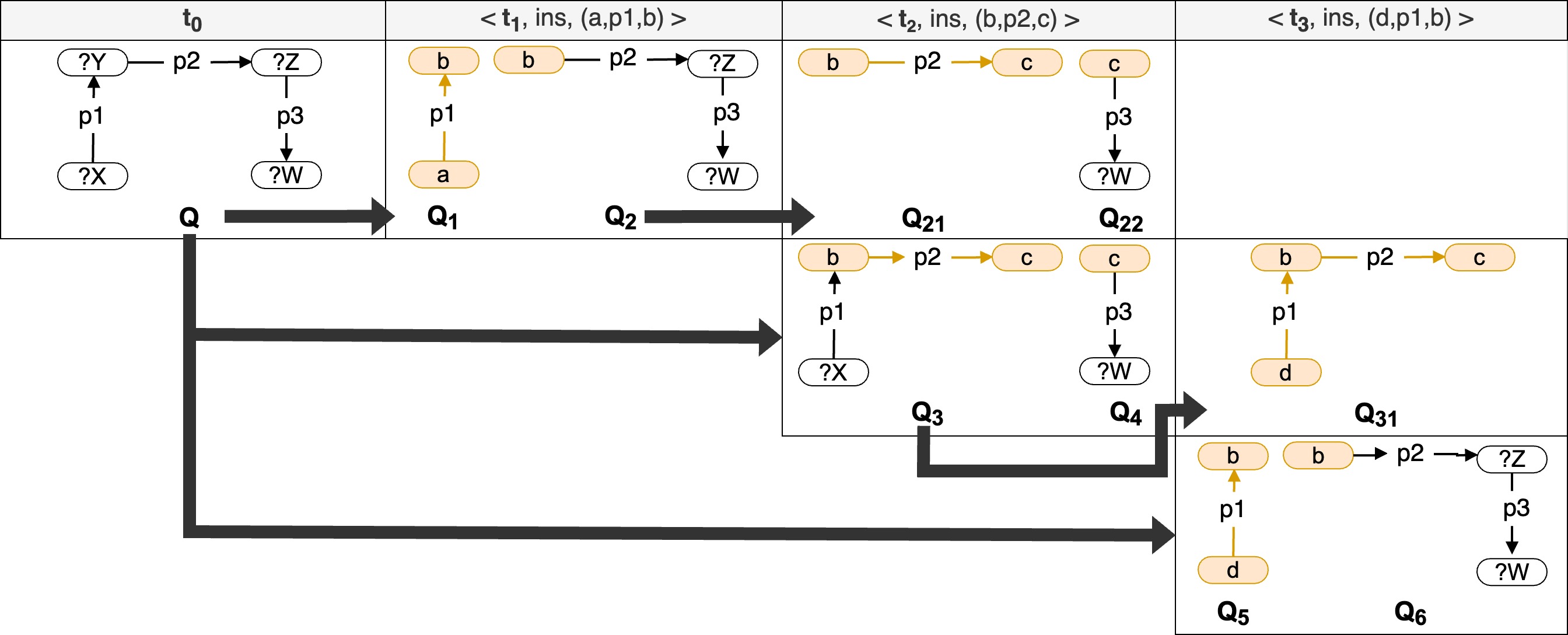}
			\caption{Var-connected queries - Recursive decomposition}
			\label{fig:connectedvarsdecomposition}
		\end{center}
	\end{figure}

In this section, we discuss how var-connected queries could be continuously evaluated.
In each triple of a  var-connected query $Q$, either the subject or the object (or both) is variable.
Hence, the first edge $e$ of the data graph $G$, which is an edge obtained through an insert update message and `matches' with a triple $r$ in $Q$,
instantiates all variables in $vars(r)$. In other words, it instantiates either a single variable or a pair of variables of $Q$ with constant values.
In this way, we get an (possibly ground) instance $Q'$ of the query $Q$. Let's refer to such queries as \textit{partial instances} of $Q$.
One could think the following approach. If $Q'$ is ground, or simple var-centric, we could proceed by evaluating $Q'$ by applying one of the algorithms presented in  Subsections~\ref{subsec:ground} or ~\ref{subsec:var-centric}, respectively.
	%
If $Q'$ is not in one of the above forms, we can apply Definition~\ref{def:cvd} to get its connected-variable decomposition ${\cal D_{CV}}(Q')$, and then apply iteratively the same evaluation strategy for all queries in ${\cal D_{CV}}(Q')$.
However, such an iterative approach is ineffective since the number of queries maintained is exponentially growing.
Fig.~\ref{fig:connectedvarsdecomposition} illustrates an example of this approach, where the arrows show the queries decomposed and instantiated at each time. As we can see,  at time $t_3$, the decomposition applied does not take into account that $Q_6$ (which equals $Q_2$) has already been decomposed into $Q_{21}$ and $Q_{22}$. Hence, we need to keep track of all the intermediate partial instances in order to ensure that we do not miss any solution; which requires high space and time complexity. In addition, we can see that the amount of queries need to be maintained is quite large, and there might be duplicate queries (e.g., $Q_2=Q_6$, $Q_{22}=Q_4$) that could increase the number of answers produced. The latter issue becomes more challenging when deletions are received.
	
\begin{figure}
\begin{center}
		\includegraphics[width=1\linewidth]{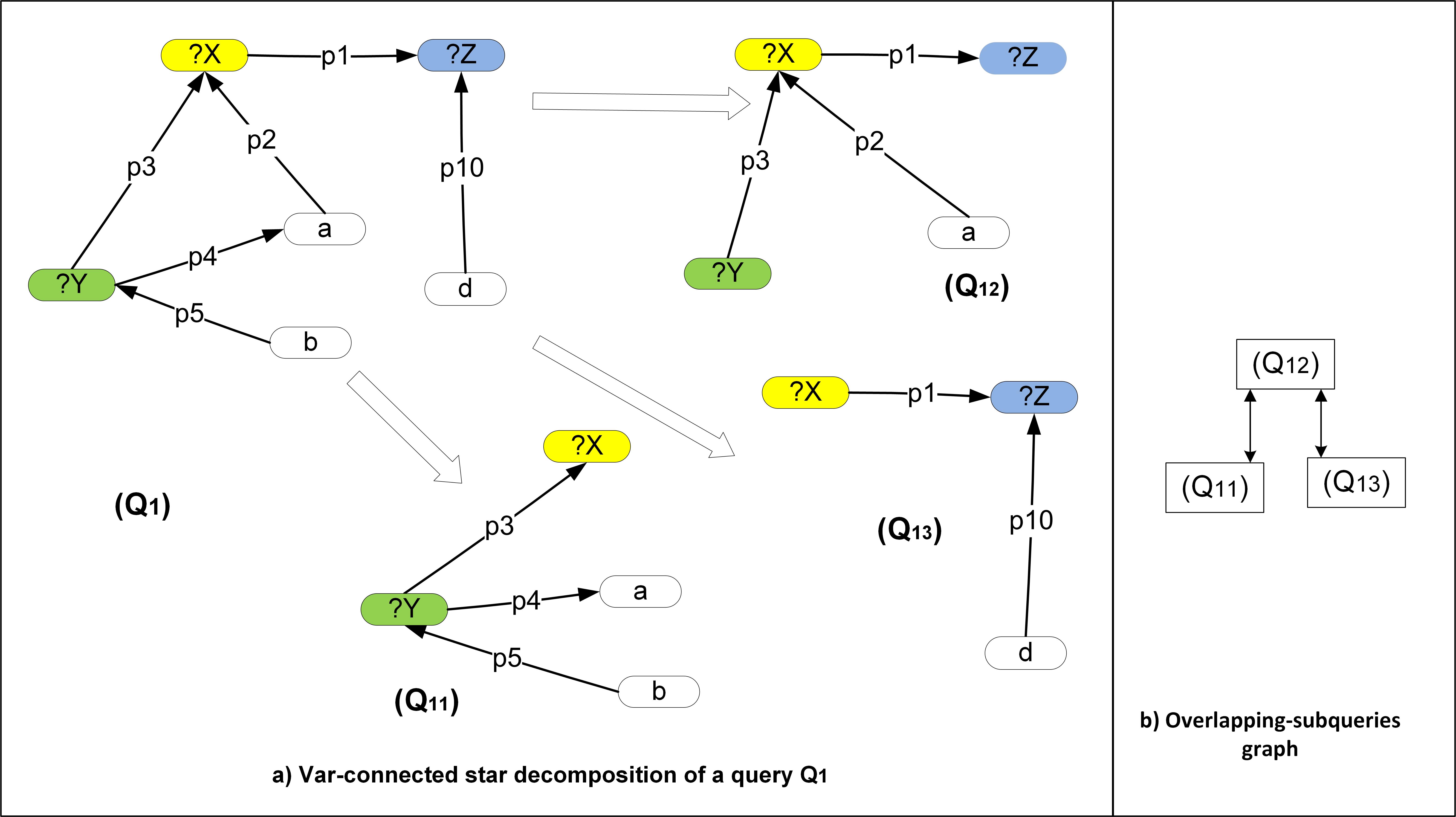}
		\caption{a) Var-connected star decomposition of a query Q1, b) Overlapping-subqueries graph.}
		\label{fig:connectedvarsstardecomposition}
\end{center}
\end{figure}

To leverage the previously-computed (partial) instances and decrease the amount of intermediate query instances, we consider a star decomposition $\CD_s(Q)$ over all the variables of the given query $Q$, so that each triple having a single variable is included in a single subquery and each subquery in $\CD_s(Q)$ is var-centric.
For example, Fig.~\ref{fig:connectedvarsstardecomposition} illustrates the star decomposition of the var-connected query $Q_{1}$ depicted in Fig.~\ref{fig:connected}. Notice that each variable in $Q_{1}$ is a central node of a subquery and each subquery in the decomposition shares at least one edge (including the corresponding variables) with another subquery into the decomposition. The following proposition generalizes this observation for any var-connected query, and is typically implied by Definitions~\ref{def:connected-variable-query}, \ref{def:stardecomposition}.

\begin{proposition}
Let $Q$ be a var-connected query and $\CD_s(Q)$ be its star decomposition. For each subquery $Q_s\in\CD_s(Q)$, there is at least one subquery $Q_s'\in\CD_s(Q)$ such that $Q_s\cap Q_s'\neq\emptyset$ and $|{\cal V}(Q_s)\cap {\cal V}(Q_s')|\geq 2$.
\end{proposition}
	
By analyzing the subqueries sharing edges, we come up with a bidirected graph, named \textit{overlapping-subqueries graph} of the decomposition, which includes a bidirected edge for each pair of queries sharing at least one edge. For example, Fig.~\ref{fig:connectedvarsstardecomposition}(b) illustrates the overlapping-subqueries graph of the star decomposition depicted in Fig.~\ref{fig:connectedvarsstardecomposition}(a). It is easy to verify that the definitions of var-connected queries and star decomposition (Definitions~\ref{def:connected-variable-query}, \ref{def:stardecomposition}) imply that the overlapping-subqueries graph of a star decomposition is always connected. This conclusion implies the following proposition, which is a valuable property used to find the delta embeddings as will discuss in the Algorithm~\ref{alg:vars-connected-queries}.

\begin{proposition}
\label{prop:array-of-subqueries}
	  	Considering a var-connected query $Q$, an edge $e$ of $Q$ and its star decomposition $\CD_s(Q)$, there is at least one (ordered) array $\CD_o=[Q_1,Q_2,\dots,Q_n]$ of the subqueries in $\CD_s(Q)$, such that
    \begin{enumerate}
	  	\item  the subqueries having as central node a variable in $\CV(e)$ take the first $|\CV(e)|$ positions in $\CD_o$, and
	  	\item for each $k=2,\dots,n$, there is a $m\leq k$ so that $Q_k\cap Q_m\neq\emptyset$.
	\end{enumerate}
\end{proposition}
	
Proposition~\ref{prop:array-of-subqueries} is easily proved by considering a graph traversal algorithm (i.e., either Breadth First Search or Depth First Search) over the overlapping-subqueries graph of $\CD_s(Q)$ and starting from the subqueries having as central node a variable in $\CV(e)$. Consider a procedure $findOrderedArray(Q,\CD_s(Q),e)$ computing such an ordered array, where  $Q$ is a var-connected query, $\CD_s(Q)$ is its star decomposition and $e$ is an edge of $Q$.
	
	To continuously find the delta solutions of a var-connected query $Q$, we aim to use an extension of the Algorithm~\ref{alg:var-centric} presented in Subsection~\ref{subsec:var-centric}. While Algorithm~\ref{alg:var-centric} accepts \textit{simple} var-centric queries, the star decomposition of $Q$ includes var-centric \textit{generalized} star queries. Hence, we extend Algorithm~\ref{alg:var-centric} to support variables not only in the central node but also in the non-central nodes. To do so, we initially define, for a query $Q$, an array $V$, called \textit{variables-summary}, whose length equals the number of variables of $Q$. The $k$-th position of $V$ includes either a value of the $k$-th variable of $Q$ or an unknown value (e.g., None - we use the symbol $'*'$ to represent this value). In the following, we use the concept of \textit{compatible} variables-summaries to describe two variables-summaries that have the same length and for each position either the corresponding values match or one of them is unknown. It is easy to see that, in essence, the variables-summaries represent partial embeddings. If two variables-summaries $V_1$, $V_2$ are compatible, we define their \textit{join}, denoted  $V_1 \otimes  V_2$, as the variables-summary constructed from $V_1$, $V_2$ as follows: for each $i\in 1,\dots,|V_1|$, we have that
	\[ (V_1 \otimes  V_2)[i] = \left\{ \begin{array}{ll}
		V_1(i) & \mbox{, if $V_1(i)\neq '*'$} \\
		V_2(i) & \mbox{, otherwise.}
	\end{array}
	\right. \]
	In this context, we extend the list $L_{Q_S}^e$ of ground var-centric instances of a star query $Q_S$ (called \textit{extended  list of ground var-centric instances}) as follows. $L_{Q_S}^e$ includes triples of the form $(X_G, Q_G, (M_{Q_G},V))$, where

\begin{enumerate}
		\item $X_G$ is an instance of the central variable $X$ of $Q_S$,
		\item $Q_G$ is the ground instance of $Q_S$ obtained by replacing (1) $X$ by  $X_G$ and (2) all the other variables of $Q_S$ by a special constant which is not included in $(U_{so} \cup L)$, and
		\item $(M_{Q_G},V)$ is a pair consisting of the triple match state $M_{Q_G}$ of $Q_G$ and a variables-summary $V$ of the variables of $Q_S$.
\end{enumerate}

	Note that although $L_{Q_S}^e$ might include multiple triples of the same instance of the central variable, it can be implemented by grouping all the triples of the same instance of the central variable to a single triple (e.g., having a set of key-lists pairs of the form $Y:[Y_1,\dots,Y_{\ell}]$ extended the variables-summary, where $Y$ is a variable and $Y_1,\dots,Y_{\ell}$ is the list of instances of $Y$). For simplicity, we use the variables-summary to describe our approach, although it has an impact on the space required to store the intermediate data.
	
	The procedure in Algorithm~\ref{alg:var-centric-ext} is the extended version of the procedure $process\_update\_message\_on\_star$ in Algorithm~\ref{alg:var-centric} and describes how an extended list of ground var-centric instances of a var-centric generalized star query is updated. Typically, it works as the $process\_update\_message\_on\_star$, but also ensures that the corresponding variables-summaries are properly updated for each update message.

		\begin{algorithm}
			\scriptsize
			\SetAlgoLined
			\SetKwFor{Loop}{Loop}{}{EndLoop}
			\SetKwInOut{Output}{output}
			
			\BlankLine
			\DontPrintSemicolon

			\textbf{Procedure} process\_update\_message\_on\_star\_ext(${\cal U}_k$, $Q_S$, $e_j$, $L^e$, $V_Q$)\;
			\PrintSemicolon
			\KwData{${\cal U}_k$: an update message $<k, Op, e>$; $Q_S$: a var-centric generalized star query; $L^e$: extended list of ground var-centric instances of $Q_S$; $V_Q$: the variables-summary of a query $Q$ having only unknown values; $e_j$: the edge of $Q_S$ unified with $e$.}
			\KwResult{A list $\CR$ of variables-summaries of all the positive/negative embeddings of $Q_S$ identified for the given update;}
			%
			
			$\CR=\emptyset$\;
			\If{$e_j$ has two variables} {
				let $X$ be the central variable of $Q_S$ and $Y$ be a non-central variable of $Q_S$\;
				let $X_G$, $Y_G$ be the values of $X$, $Y\in{\cal V}(Q_S)$, respectively, obtained by the unification of $e$ with $e_j$\;
			}
			let $Q_G$ be the instance of $Q_S$ obtained by replacing $X$ by  $X_G$ and all the other variables of $Q_S$ by a special constant\;
			\eIf{$Op = ins$}  {
					\ForEach{triple $(X_c, Q_c, (M_{Q_G},V)) \in L^e$ s.t. $X_c=X_G$ and $Q_c=Q_G$}  {
						\eIf{$e_j$ has two variables} {
							\eIf{$V[idx(Y)]\neq '*'$} {
								Create a copy $(X_G, Q_G, (M_{Q_G}',V'))$ of $(X_G, Q_G, (M_{Q_G},V))$\;
								$M_{Q_G}'[j] =  1$\;
								\tcc{$idx(Y)$ represents the position of $Y$ in $V'$}
								$V'[idx(Y)]=Y_G$\;
								$L^e = L^e \cup \{(X_G, Q_G, (M_{Q_G}',V'))\}$\;
								\lIf{ground\_BGP\_eval($M_{Q_G}'$, $Q_G$) = true} {$\CR=\CR\cup\{V'\}$}
							}
							{
								$M_{Q_G}[j] =  1$\;
								\tcc{$idx(Y)$ represents the position of $Y$ in $V$}
								$V[idx(Y)]=Y_G$\;
								\lIf{ground\_BGP\_eval($M_{Q_G}$, $Q_G$) = true} {$\CR=\CR\cup\{V\}$}
							}
						}{
							$M_{Q_G}[j] =  1$\;
							\lIf{ground\_BGP\_eval($M_{Q_G}$, $Q_G$) = true} {$\CR=\CR\cup\{V\}$}
						}
					}
					\If{there is no triple $(X_G, Q_G, (M_{Q_G},V)) \in L^e$, for some $(M_{Q_G},V)$}{
						\lFor{$i = 1$  \KwTo  $|Q_S|$} {$M_{Q_G}[i] = 0$}
						$M_{Q_G}[j] = 1$\;
						Create a copy $V$ of $V_Q$\;
						$V[idx(X)]=X_G$\;
						\lIf{$e_j$ has two variables} {
							$V[idx(Y)]=Y_G$
						}
						$L^e = L^e \cup \{(X_G, Q_G, (M_{Q_G},V))\}$\;
						\lIf{$|Q_S| = 1$} {$\CR=\CR\cup\{V\}$}
					}
			}
			{
						\ForEach{$m = (X_G, Q_G, (S''_{Q'_G},V'')) \in L^e$ s.t. $S''_{Q_G}[j] = 1$} {
							\lIf{ground\_BGP\_eval($S''_{Q_G}$, $Q_G$) = true}{$\CR=\CR\cup\{V''\}$}
								$S'_{Q_G}[j] = 0$\;
								\lIf{$e_j$ has two variables} {
									$V''[idx(Y)]='*'$
								}
								\tcp{Once we apply a deletion of an edge, we might have multiple identical records in $L^e$. In such a case, we remove the last one produced.}
								\If{all\_zero($S''_{Q_G}, Q_G$) = true or there is $t\in L^e$ s.t., the variables-summaries of $t$ and $m$ are compatible} {Remove $m$ from $L^e$}
							}
			}
			\KwRet{$\CR$}
			\caption{Algorithm updating extended lists of  var-centric star queries}
			\label{alg:var-centric-ext}
		\end{algorithm}

	Let us, now, present the main algorithm illustrated in Algorithm~\ref{alg:vars-connected-queries} for continuously computing var-connected queries. The main idea behind this algorithm is to recursively check the extended lists of the star queries in the decomposition of the var-connected query. Both insertion and deletion are handled similarly. In particular, each edge $e$ received by the update stream that unifies with an edge $e_j$ of the var-connected query $Q_{vc}$ typically instantiates the variable(s) of $e_j$; one or two variables. If $e_j$ includes a single variable, then a single subquery in the star decomposition $\CD_s(Q_{vc})$ of $Q_{vc}$ is updated (regardless of the update operation); specifically, the query having as central node this variable. Similarly, if $e_j$ includes two variables, both the subqueries with central nodes these variables should be updated. Note here that although the edge $e$ offers a single instance of the central variable, there might be multiple records in the extended list of the corresponding subquery, since each subquery includes at least two variables (one as central node and one or more adjacent to it). For example, if we have already received two messages inserting edges that match the edge $(?Y,p3,?X)$ and have the same instance $Y_G$ for $?Y$, we will have two triples in $L^e_{Q_{11}}$. Both these triples should be updated once we receive a message inserting the triple $(Y_G,p4,a)$. Hence, each update message ${\cal U}_k$ determines an update operation over one or two extended lists, according to the number of variables included in $e_j$.
	
	Now, if each of these updates makes a single subquery included in those lists ground (in case of insertion), then such an operation triggers several checks that are required in order to identify whether a positive answer is constructed. In essence, we just need to search all the ground instances of the remaining star subqueries in order to construct a positive answer, if there is any. Similarly, if the operation specified in ${\cal U}_k$ is deletion and we delete the given edge from a ground instance, then we need to check whether a negative answer is constructed. To do so, we use the overlapping-subqueries graph to propagate the checking task to the adjacent subqueries, recursively, and for each ground subquery found in each list. If all the subqueries have a ground instance connected with the ground query affected, then either a positive or negative answer is constructed and returned by the algorithm. Note here that using the overlapping-subqueries graph we can dynamically order the subqueries in the decomposition in order to ensure that we start with the subqueries affected and propagate the check into the remaining ones.
	
		\begin{algorithm}
		\caption{Algorithm that Evaluates var-connected BGP queries}
		\label{alg:vars-connected-queries}
		\scriptsize
		\SetAlgoLined
		\DontPrintSemicolon
		\SetKwFor{Loop}{Loop}{}{EndLoop}
		\SetKwInOut{Output}{output}
		
		\textbf{Procedure} eval\_vars\_connected\_query($Q_{vc}$)\;
		\PrintSemicolon
		\KwData{$Q_{vc}$: a var-connected query;}
		\KwResult{Positive/negative answers;}
		
		Construct a star decomposition $\CD_s(Q_{vc})$ of $Q_{vc}$\;
		Construct the variables-summary of $Q_{vc}$ and initialize it with unknown values\;
		\ForEach{variable $X$ of $Q_{vc}$ which is the central node of $Q\in\CD_s(Q_{vc})$} {
			Construct an empty extended list $L^e_{Q}$ of ground var-centric instances of $Q$\;
		}
		\Loop{}
		{  Get the next update message ${\cal U}_k =  <k, Op, e>$\;
			\eIf{$e$ does not unify with any edge in $Q_{vc}$}{
				\textbf{continue}\;
			}
			{
				\ForEach{$e_j$ in $Q_{vc}$, with $1 \leq j \leq |Q_{vc}|$, s.t. $e$ unifies with $e_j$} {
					$\CD_o = findOrderedArray(Q_{vc},\CD_s(Q_{vc}),e_j)$\;
					$Q_1=\CD_o[0]$\;
					$R_1$ = $process\_update\_message\_on\_star\_ext$(${\cal U}_k$, $Q_1$, $e_j$, $L^e_{Q_1}$, $V_{Q_{vc}}$)\;
					$R=\emptyset$\;
					\eIf{$|\CV(e_j)|=2$}{
						\tcp{i.e., $e_j$ includes two variables}
						$Q_2=\CD_o[1]$\;
						$R_2$ = $process\_update\_message\_on\_star\_ext$(${\cal U}_k$, $Q_2$, $e_j$, $L^e_{Q_2}$, $V_{Q_{vc}}$)\;
						\ForEach{$V_1\in R_1$ and $V_2\in R_2$ s.t. $V_1, V_2$ are compatible}{
							$R=R\cup\{V_1 \otimes  V_2\}$\;
						}
					}
					{
						$R=R_1$\;
					}
					\ForEach{$V_k\in R$}{
						$\CV_{tmp}=$find\_delta\_var\_summaries($V_k$, $|\CV(e_j)|$, $D_o$)\;
						\ForEach{$V_t\in\CV_{tmp}$}{
							let $Q_{vc}^G$ be the query constructed by replacing the variables in $\CV(Q_{vc})$ with the corrseponding values in  $V_t$\;
							\eIf{$Op=ins$}
							{
								\Output{$< Q_{vc}^G, positive, V_t >$}
							}
							{\tcp{Case where $Op=del$}
								\Output{$< Q_{vc}^G, negative, V_t >$}
							}
						}
					}
					
				}
			}
			
		}
	\end{algorithm}
	
	Typically, the procedure $process\_update\_message\_on\_star\_ext$ described in Algorithm~\ref{alg:var-centric-ext} performs the update of the extended lists of the subqueries with central nodes in $\CV(e_j)$. The procedure $find\_delta\_var\_summaries$ presented in Algorithm~\ref{alg:vars-connected-find-solutions} recursively checks whether an extended list of a subquery has a ground instance affected by the updated operation, and returns the corresponding variables-summaries, if there is any, in order to be merged and the delta answers to be properly constructed. Note that, initially, the procedure $find\_delta\_var\_summaries$ checks whether there is any ground instance (see in line 5 the function $getGroundItems$) in the extended list. If there is no ground instance, typically there is no delta answer produced, hence, there is no need to check the upcoming lists. Furthermore, it is easy to verify that there are two cases where a record in an extended list needs to be removed. Both cases can appear when ${\cal U}_k$ requests a deletion operation. One case is when a record becomes completely empty (i.e., each position of the triple match state is $0$). Secondly, a deletion might imply multiple records with compatible variables-summaries. In such a case, if we do not ensure that a single record is included in the extended list, we might have duplicate delta answers; hence, we remove the one produced after the deletion.

\begin{algorithm}
	\caption{Recursive function finding delta answers}
	\label{alg:vars-connected-find-solutions}
	\scriptsize
	\SetAlgoLined
	\DontPrintSemicolon
	\SetKwFor{Loop}{Loop}{}{EndLoop}
	\SetKwInOut{Output}{output}
	
	\textbf{Function} find\_delta\_var\_summaries(variables-summary $V_i$, index $k$, array of subqueries $D_o$)\;
	\PrintSemicolon
	\lIf{$k>(|D_o|-1)$}{
		\KwRet{$(True, \{V_i\})$}
	}
	$Q_i=D_o[k]$\;
	let $C_i$ be the value of the central node of $Q_i$ in $V_i$\;
	\tcp{Returns all the variables-summaries of the triples $(X_G, Q_G, (M_{Q_G},V))$ s.t. $X_G=C_i$, $M_{Q_G}$ is complete and $V$, $V_i$ are compatible.}
	$G=getGroundItems(L^e_{Q_i}, C_i, V_i)$\;
	\If{$G\neq\emptyset$}{
		$R=\emptyset$\;
		\ForEach{$V_j\in G$}{
			$V_{ij} = V_i\otimes V_j$\;
			\eIf{$k\neq(|D_o|-1)$}{
				$R_{tmp}=$find\_delta\_var\_summaries($V_{ij}$, $(k+1)$, $D_o$)\;
				\lIf{$R_{tmp}[0]=True$}{
					$R=R\cup R_{tmp}[1]$
				}
			}{
				\KwRet{$(True, \{V_{ij}\})$}
			}
		}
		\lIf{$R\neq\emptyset$}{
			\KwRet{$(True, R)$}
		}
	}
	\KwRet{$(False, \emptyset)$} \;
\end{algorithm}

\begin{figure}[htb]
\begin{center}
\includegraphics[width=1.0\linewidth]{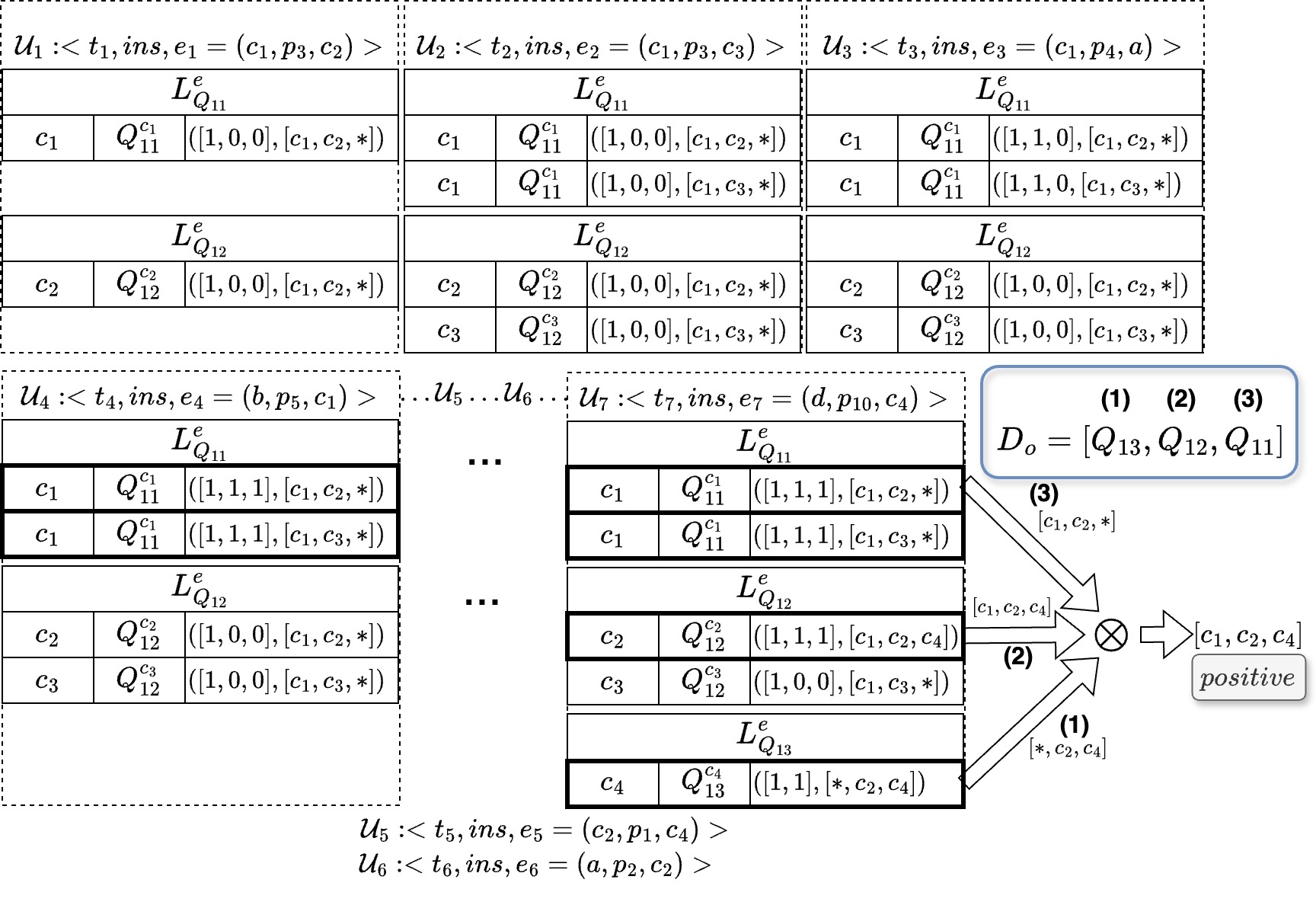}
\caption{Example - Evaluation of a var-connected query (Insertion).}
\label{fig:example-var-connected-query-evaluation}
\end{center}
\end{figure}

\begin{example}
\label{example:var-connected-queries-evaluation}
Let $Q_1$ be the var-connected query, and $\CD_s(Q_1)=\{Q_{11},Q_{12},Q_{13}\}$ be its star decomposition  (line 4, Algorithm~\ref{alg:vars-connected-queries}) depicted in Fig.~\ref{fig:connectedvarsstardecomposition}(a). Initially, the variables-summary $V=[*,*,*]$  of $Q_1$ is constructed (line 5, Algorithm~\ref{alg:vars-connected-queries}), where $'*'$ represents an unknown value. Table~\ref{tbl:mapping-triple-match-state-and-var-summary} shows the mapping between the variables of $Q_1$ and the positions of the array $V$. We then construct an empty extended list of ground var-centric instances for each query in $\CD_s(Q_1)$; i.e., $L^e_{Q_{11}}$, $L^e_{Q_{12}}$, $L^e_{Q_{13}}$ for $Q_{11}$, $Q_{12}$, $Q_{13}$, respectively (lines 4-5, Algorithm~\ref{alg:vars-connected-queries}).
	
Let's now see how the Algorithm~\ref{alg:vars-connected-queries} processes each update message $\CU_k$ received (lines 7-28). In particular, consider the messages $\CU_k=<t_k,ins,e_k>$ received at times $t_k$, with $k=1,\dots,7$, where the edges $e_k$ are depicted in Fig.~\ref{fig:example-var-connected-query-evaluation}.
Furthermore, Fig.~\ref{fig:example-var-connected-query-evaluation} illustrates the status of each extended list of ground var-centric instances after processing each message $\CU_k$, where each dashed box shows the snapshot of each list after processing the corresponding edge at each time (for reducing the image space we avoid depicting the snapshot of each list at time $t_5$ and $t_6$).
	 At time $t_1$ (i.e., $k=1$), we receive a message requesting an insertion of the triple $e_1=(c_1,p_3,c_2)$. Checking whether $e_1$ unifies with an edge in $Q_1$ (line 8), we conclude that it unifies with the query edge $e_j=(?Y,p_3,?X)$. Hence, we initially compute an ordered array $D_o=[Q_{11},Q_{12},Q_{13}]$ (line 12) based on the overlapping-subqueries graph $G_{Q_1}$ of $\CD_s(Q_1)$ depicted in Fig.~\ref{fig:connectedvarsstardecomposition}(b). Note that the ordered array $D_o$ is typically computed for each message received, however, it can be precomputed for each subquery and cached to improve the performance of the algorithm in each iteration.
	Since $e_j$ has two variables, we update two lists; specifically, the lists $L^e_{Q_{11}}$ and $L^e_{Q_{12}}$ of the first two subqueries in $D_o$ (lines 14 and 18), i.e., the subqueries $Q_{11},Q_{12}$ which have as central node a variable in $e_j$ (i.e., $?Y,?X$, respectively).
Hence, we call the procedure defined in Algorithm~\ref{alg:var-centric-ext} for both $Q_{11}$ and $Q_{12}$ to update the lists $L^e_{Q_{11}}$, $L^e_{Q_{12}}$.
	
The Algorithm~\ref{alg:var-centric-ext} (A\ref{alg:var-centric-ext}, for short) operates on $L^e_{Q_{11}}$, as follows. Initially, we contruct an instance $Q_G$ of $Q_{11}$ s.t. $?Y=c_1$ and $?X=c_2$; hence $Q_G=\{(c_1,p_3,c_2),(c_1,p_4,a),(c_1,p_5,b)\}$ (line 6). Since $L^e_{Q_{11}}$ is empty (lines 8-21 are ignored), we create the first triple of $L^e_{Q_{11}}$ by initially constructing its triple match states (lines 24-25 of A\ref{alg:var-centric-ext}). Consider the mapping between $Q_{11}$'s edges and the positions of $M_{Q_{11}}$ array (see Table~\ref{tbl:mapping-triple-match-state-and-var-summary}). Since $e_1$-edge unifies with $e_j$, which is mapped to the first position of $M_{Q_{11}}$, we have that $M_{Q_G} =[1,0,0]$. $M_{Q_G}$ represents that an instance of the edge mapped in the first position has been inserted. A variables-summary is then constructed from $V$ by adding the values of $?X$ and $?Y$ to the corresponding positions (lines 26-28 of A\ref{alg:var-centric-ext}), taking into account the mapping defined in Table~\ref{tbl:mapping-triple-match-state-and-var-summary}; i.e., $V=[c_1,c_2,*]$. To complete the update of $L^e_{Q_{11}}$ we insert the record $(c_1,Q_{11}^{c_1},([1,0,0],[c_1,c_2,*]))$ to this list, where $Q_{11}^{c_1}=Q_G$. Similarly, calling the Algorithm~\ref{alg:var-centric-ext} for $Q_{12}$ and $e_1$ (lines 17-19), we insert the record $(c_2,Q_{12}^{c_2},([1,0,0],[c_1,c_2,*]))$ to $L^e_{Q_{12}}$. After processing the message $\CU_1$, the lists' statuses are illustrated in the first box of Fig.~\ref{fig:example-var-connected-query-evaluation}. Note that since $c_1$ is the value of the central variable $?Y$, it is added in the first column and it is typically used as a key to a Hash Map structure. Moreover, since $L^e_{Q_{13}}$ is still empty at this time, we omit to display it in Fig.~\ref{fig:example-var-connected-query-evaluation}.
	
	At time $t_2$, the insertion of $e_2$ is requested through the message $\CU_2$. Notice that $e_2$ unifies with the same edge $e_j$ of $Q_{11}$ and $Q_{12}$; hence, only $L^e_{Q_{11}}$ and $L^e_{Q_{12}}$ are affected and need to be updated (see the second box in Fig.~\ref{fig:example-var-connected-query-evaluation}). The difference, now, with the previous iteration is that both $L^e_{Q_{11}}$ and $L^e_{Q_{12}}$ are not empty. Hence, lines 8-22 of A\ref{alg:var-centric-ext} are triggered (instead of the lines 23-30 applied in the previous iteration). Since $e_2$ provides a new instance of the same variables (i.e., $V[idx(?Y)]=V[idx(?X)]=1$, line 10 of A\ref{alg:var-centric-ext}), we end up with a new record in  $L^e_{Q_{11}}$  (resp. $L^e_{Q_{12}}$) which is $(c_1,Q_{11}^{c_1},([1,0,0],[c_1,c_3,*]))$ (resp. $(c_3,Q_{12}^{c_3},([1,0,0],[c_1,c_3,*]))$), through the steps described in lines 11-13. To complete the procedure, we check (line 15 of A\ref{alg:var-centric-ext}) whether the triple match state of the currently-created record is complete (i.e., it has the value $1$ at each position). The triple match state is still not complete, and the output $\CR$ of the procedure remains empty.
	
	Moving to the message $\CU_3$ and following the same steps, only the list $L^e_{Q_{11}}$ is updated (see the third box in Fig.~\ref{fig:example-var-connected-query-evaluation}); $e_3$ unifies, only, with the $Q_{11}$'s edge $(?Y,p_4,a)$, which includes a single variable. In the processing step described by lines 20-22 of A\ref{alg:var-centric-ext}, we update the triple match state by adding the value 1 to its second position (according to the mapping described in Table~\ref{tbl:mapping-triple-match-state-and-var-summary}) and checking whether we have a complete triple match state. Both records in $L^e_{Q_{11}}$ have the same central node (the condition in line 8 of A\ref{alg:var-centric-ext} is satisfied for both records of $L^e_{Q_{11}}$), and they are updated as previously described.
	
	The same steps are also followed at time $t_4$, but in this case, we have a complete triple match state for both records of $L^e_{Q_{11}}$ (see the fourth box in Fig.~\ref{fig:example-var-connected-query-evaluation}, where the complete records are highlighted through thick borders). The condition in line 22 of A\ref{alg:var-centric-ext} is satisfied, and the output of the procedure includes the instance of variables-summaries for both records; i.e., $\CR=\{[c_1,c_2,*],[c_1,c_3,*]\}$. As a consequence of the non-empty output, the steps described by lines 21-28 of Algorithm~\ref{alg:vars-connected-queries} are triggered.
	
	Lines 23-28 of Algorithm~\ref{alg:vars-connected-queries} construct the delta answers, if exist. At time $t_4$, the procedure described by the Algorithm~\ref{alg:vars-connected-find-solutions} is called (line 24, Algorithm~\ref{alg:vars-connected-queries}), for each $V_k\in{\CR}$, where the index $k$ equals $1$. In Algorithm~\ref{alg:vars-connected-find-solutions}, we check whether there are complete triple match states of other subqueries that can be combined and produce delta answers. To do so, Algorithm~\ref{alg:vars-connected-find-solutions} (A\ref{alg:vars-connected-find-solutions}, for short) operates recursively following the order of the subqueries in $D_o$ (line 11, A\ref{alg:vars-connected-find-solutions}). In particular, for the variables-summary $V_i=[c_1,c_2,*]\in\CR$, we initially check whether there is a record in the extended list $L^e_{Q_{12}}$ (since $Q_i=Q_{12}=D_o[1]$, line 3 of A\ref{alg:vars-connected-find-solutions}) that has a complete triple match state and its central node equals $c_1$ (line 5, A\ref{alg:vars-connected-find-solutions}). As we can easily verify, there is no such record at this time. We get the same result for the second variables-summary in $\CR$, hence, we do not have any delta answer produced by processing $\CU_4$.
	
	Algorithm~\ref{alg:vars-connected-queries} operates similarly for the messages $\CU_5$ and $\CU_6$, where no delta answer is resulted. At time $t_7$, however, we are able to find a delta answer as follows. We can see that after updating the lists at time $t_7$, we have three records with complete triple match state, which are also compatible with each other and can be joined in order to give the positive answer $[c_1,c_2,c_4]$ (see the last box in Fig.~\ref{fig:example-var-connected-query-evaluation}).
	
	Let's see now how this answer is computed by the algorithm. $\CU_7$ message typically requests the insertion of the edge $e_7=(d,p_{10},c_4)$, which unifies with the query edge $e_j=(d,p_{10},?Z)$ in $Q_{13}$. The updated extended list $L^e_{Q_{13}}$ resulted in line 14 of Algorithm~\ref{alg:vars-connected-queries} includes a single variables-summary, i.e., $[*,c_2,c_4]$. Next, Algorithm~\ref{alg:vars-connected-find-solutions} is triggered in line 24 of Algorithm~\ref{alg:vars-connected-queries}, for $V_k=[*,c_2,c_4]$ and index $k=1$. The $D_o$ constructed at this iteration is given by the array $[Q_{13},Q_{12},Q_{11}]$, which is also illustrated in Fig.~\ref{fig:example-var-connected-query-evaluation}. Algorithm~\ref{alg:vars-connected-find-solutions}, then, operates as follows. Initially, it checks for compatible variables-summaries in $L^e_{Q_{12}}$ (line 5, A\ref{alg:vars-connected-find-solutions}), since $Q_i=D_o[1]=Q_{12}$ (line 3, A\ref{alg:vars-connected-find-solutions}). The function $getGroundItems$  searches $L^e_{Q_{12}}$ for all the records having a complete triple match state and a node labeled by $C_i=c_2$ as the central node. Its result is $G=\{[c_1,c_2,c_4]\}$. Then, $[c_1,c_2,c_4]$ and $[*,c_2,c_4]$ are joined, resulting the variables-summary $V_{ij}=[c_1,c_2,c_4]$ (line 9, A\ref{alg:vars-connected-find-solutions}). Since there is one more subquery that needs to be checked in order to identify a delta answer (i.e., $k=1\neq (|D_o|-1)=2$, line 10 of A\ref{alg:vars-connected-find-solutions}), we call the same procedure for the combined variables-summary $V_{ij}$ and the next subquery in $D_o$ (i.e., $D_o[k+1]=D_o[2]=Q_{11}$, line 11 of A\ref{alg:vars-connected-find-solutions}). At this iteration, A\ref{alg:vars-connected-find-solutions} computes a new $V_{ij}=[c_1,c_2,c_4]$ by joining the previously-computed $V_{ij}$ and the variables-summaries $[*,c_2,c_4]$ found in $L^e_{Q_{11}}$. The combined variables-summaries are added into the output $\CR$ ( line 12 of A\ref{alg:vars-connected-find-solutions}), duplicates are removed, and the result is returned to Algorithm~\ref{alg:vars-connected-find-solutions}, which then constructs the delta answers as described in lines 25-28. Fig.~\ref{fig:example-var-connected-query-evaluation} illustrates the recursive joining process, where the numbers $(1)-(3)$ represent the order of the nested join, according to the order defined in $D_o$.
\end{example}

\begin{center}
\begin{table}[htb]
\begin{tabular}{|cccc|cc|}
	\hline
	\multicolumn{4}{|c}{\textbf{Triple match state}} & \multicolumn{2}{|c|}{\textbf{Variables-summaries}}\\
	\hline
		$M_{Q_{1i}}$ & $1^{st}$ position & $2^{nd}$ position & $3^{rd}$ position & Variable & Position\\\hline
		$M_{Q_{11}}$ & $(?Y,p_3,?X)$ & $(?Y,p_4,a)$ & $(?Y,p_5,b)$ & $?Y$ & $1$\\
		$M_{Q_{12}}$ & $(?Y,p_3,?X)$ & $(a,p_2,?X)$ & $(?X,p_1,?Z)$ & $?X$ & $2$\\
		$M_{Q_{13}}$ & $(?X,p_1,?Z)$ & $(d,p_{10},?Z)$ &  & $?Z$ & $3$\\
	\hline
\end{tabular}
\caption{Example~\ref{example:var-connected-queries-evaluation} - Triple match states and variables-summaries}
\label{tbl:mapping-triple-match-state-and-var-summary}

		\end{table}
	\end{center}
	
The following proposition shows the time and space complexity of the Algorithm~\ref{alg:vars-connected-queries}.
	
\begin{proposition}
\label{prop:vars-connected-complexity}
Algorithm~\ref{alg:vars-connected-queries} requires polynomial space and time to compute delta embeddings of an input message.
\end{proposition}
\proof
Let $Q_{vc}$ be a vars-connected query of $v=|\CV(Q_{vc})|$ variables, $\CU_{\ell}=<t_{\ell},Op,e>$ be an update message received
at time $t_{\ell}$,
and $G=(G_0,G_1,\dots,G_{\ell})$ be the dynamic graph at time $t_{\ell}$. Assume
${\cal D_{CV}}(Q_{vc})=\{Q_1,Q_2,\dots ,Q_m\}$ is the star decomposition of $Q_{vc}$. Without loss of generality, we assume that $\CD_o=[Q_1,Q_2,\dots,Q_m]$ is an ordered set of  ${\cal D_{CV}}(Q_{vc})$ that is produced from overlapping-subqueries graph $S_G$ of  ${\cal D_{CV}}(Q_{vc})$ in order to manage the update message $\CU_{\ell}$. For each update, we need to re-order  the queries in the decomposition by parsing $S_G$, i.e.,  $O(|\CN(S_G)|+|S_G|)$ time; which is constant in terms of the data graph. Furthermore, we consider that $n$ edges have been received till $t_{\ell}$.
	
	To compute the space required to store the extended list $L^e_{Q_k}$ of ground var-centric instances of the query $Q_k$ in ${\cal D_{CV}}(Q_{vc})$, we have that $L^e_{Q_k}$ contains at most $n_{k}$ triples, one triple for each possible combination of variables; i.e., it requires $O(n^v)$ space. Hence, to store all the $m$ lists we need $O(mn^v)$ space, which is polynomial.
	
	Let us, now, focus on the time complexity. Note that all the operations over the variables-summaries and updates of each triple are performed in constant time since they depend on the number of variables and triples of the query. Suppose that at time $t_{\ell}$, we have that $n\leq n_1+\dots+n_m$, where $n_i$ is the number of edges matching an edge in $Q_i\in\CD_o$. To update the list of $Q_1$ (and $Q_2$ in case $e$ unifies with a two-variable edge of $Q_{vc}$), we need $O(n_1^{v'})$ time (procedure $process\_update\_message\_on\_star\_ext$), where $v'$ is the max number of edges of a subquery in $\CD_o$. If $e$ unifies with a two-variable edge of $Q_{vc}$, then we additionally need $O(n_2^{v'})$ time to update $Q_2$'s list and $O(n_1^{v'}n_2^{v'})$ to merge the corresponding variables-summaries. Note here that if $Op=del$ we additionally need $O(n_1^{v'})$ time (or $O(n_1^{v'}+n_2^{v'})$) to check for duplicate variables summaries in the list (lines 37-38 in Algorithm~\ref{alg:var-centric-ext}).
	Then, for the remaining subqueries, we check the ground ones and merge their variables-summaries, hence, for the subquery $Q_i\in\CD_o$, we need $O(n_1^{v'}n_2^{v'}\dots n_i^{v'})$ time; i.e., $O(n^{iv})$ time. Thus, the overall time required to update the lists and find the delta answers is $O(n^{mv})$ (i.e., polynomial).
\QED
	
	As a consequence of the Propositions~\ref{prop:ground-complexity}, \ref{prop:var-centric-complexity} and \ref{prop:vars-connected-complexity}, and the generic algorithm presented in Section~\ref{subsec:generic}, it is easy to verify that the time and space complexities of evaluating BGP queries are dominated by the evaluation of the var-connected queries. Notice that although a Cartesian product of the results of the subqueries is required, the number of subqueries is given by a constant number for each BGP query. The following proposition shows the complexity result for BGP queries, and Table~\ref{table:complexities-table} summarizes the complexity results of this work.
	
\begin{proposition}
\label{prop:BGP-complexity}
To compute the delta embeddings of a BGP query for an input message over a dynamic data graph requires polynomial space and time.
\end{proposition}

\begin{table}[htbp]\footnotesize
		\centering
		\begin{tabular}{|l|l|l|c|}
			\hline
			\multicolumn{ 1}{|c|}{\textbf{Query Classes}} & \multicolumn{ 1}{|c|}{\textbf{Time Complexity}} & \multicolumn{ 1}{|c|}{\textbf{Space Complexity}} & \multicolumn{ 1}{|c|}{\textbf{Proposition}} \\ \hline
Ground queries & $O(1)$ & $O(|Q_G|)$ & \ref{prop:ground-complexity} \\\hline
Simple var-centric queries & $O(2|Q_S|)$ & $O(n)$& \ref{prop:var-centric-complexity}\\\hline
Loosely-connected queries & $O(n+2m|Q_S|)$ & $O(mn+|Q_G|)$& \ref{prop:loosely-connected-complexity}\\\hline
Var-connected queries & polynomial & polynomial& \ref{prop:vars-connected-complexity}\\\hline
BGP queries & polynomial & polynomial& \ref{prop:BGP-complexity}\\\hline
		\end{tabular}
		\caption{Space and Time Complexity per Query type}
		\label{table:complexities-table}
	\end{table}

\section{Conclusion}
\label{sec:conclusion}
	In this work, we investigate the problem of evaluating Basic Graph Patterns (BGPs) over dynamic Linked Data graphs. In this context, we present an algorithm that computes the delta embeddings in polynomial time and requires polynomial space. Furthermore, we study the problem for a selected set of BGP subclasses, including ground BGP queries, simple var-centric star queries and loosely-connected BGP queries where we achieve significant performance improvements. The algorithms handle both insertions and deletions of triples and continuously result in delta answers as update messages are processed. The query evaluation approaches presented by the aforementioned algorithms on based on effective query decomposition to improve performance and minimize cached data.

	Both ground BGP and simple var-centric star queries are continuously evaluated by maintaining the minimum amount of data required in order to be able to efficiently find delta answers. The continuous evaluation of loosely-connected queries combines the evaluation patterns of ground BGP and simple var-centric star queries. However, evaluating var-connected queries is far more complex, since the amount of data required to be maintained is high. In future steps, we plan to improve the continuous evaluation of var-connected queries, as well as study the problem in a distributed environment and leverage computational statistics and machine learning approaches to improve the overall evaluation time.






\nocite{*}
\bibliographystyle{elsarticle-num}            
\bibliography{bibliography}        


%

\end{document}